\documentclass{IEEEtran}
\usepackage{cite}
\usepackage{amsmath,amssymb,amsfonts}
\usepackage{graphicx}
\usepackage{textcomp}
\def\BibTeX{{\rm B\kern-.05em{\sc i\kern-.025em b}\kern-.08em
    T\kern-.1667em\lower.7ex\hbox{E}\kern-.125emX}}

\usepackage{caption}
\usepackage{subcaption}
\usepackage{braket}
\usepackage{algorithm}
\usepackage{algpseudocode}
\usepackage{fancyvrb}

\begin{document}

\title{A Quantum Computer Amenable Sparse Matrix Equation Solver}
\author{
        \IEEEauthorblockN{Christopher D. Phillips\IEEEauthorrefmark{1}, \emph{Student Member, IEEE} and
                          Vladimir I. Okhmatovski\IEEEauthorrefmark{1}, \emph{Member, IEEE}}
        \\
        \IEEEauthorblockA{\IEEEauthorrefmark{1}Department of Electrical and Computer Engineering, University of Manitoba, Winnipeg, MB  R3T 2N2, Canada}
       }

\maketitle

\begin{abstract}
Quantum computation offers a promising alternative to classical computing methods in many areas of numerical science,
with algorithms that make use of the unique way in which quantum computers store and manipulate data often achieving dramatic improvements in performance over their classical counterparts.
The potential efficiency of quantum computers is particularly important for numerical simulations, where the capabilities of classical computing systems are often insufficient for the analysis of real-world problems.
In this work, we study problems involving the solution of matrix equations, for which there currently exists no efficient, general quantum procedure.
We develop a generalization of the Harrow/Hassidim/Lloyd algorithm by providing an alternative unitary for eigenphase estimation.
This unitary, which we have adopted from research in the area of quantum walks, has the advantage of being well defined for any arbitrary matrix equation, 
thereby allowing the solution procedure to be directly implemented on quantum hardware for any well-conditioned system.
The procedure is most useful for sparse matrix equations, as it allows for the inverse of a matrix to be applied with $\mathcal{O}\left(N_{nz}\log\left(N\right)\right)$ complexity,
where $N$ is the number of unknowns, and $N_{nz}$ is the total number of nonzero elements in the system matrix.
This efficiency is independent of the matrix structure, and hence the quantum procedure can outperform classical methods for many common system types.
We show this using the example of sparse approximate inverse (SPAI) preconditioning, which involves the application of matrix inverses for matrices with $N_{nz}=\mathcal{O}\left(N\right)$.
While these matrices are indeed sparse, it is often found that their inverses are quite dense, and classical methods can require as much as $\mathcal{O}\left(N^3\right)$ time to apply an inverse preconditioner.
\end{abstract}

\markboth{This work has been submitted to the IEEE for possible publication. Copyright may be transferred without notice.}{}


\section{Introduction}\label{sec:intro}
Matrix equation formulations are a powerful tool for the numerical analysis of physical systems, and their usage is widespread across many areas of engineering \cite{Chandrupatla,Gibson,Colaco}.
Unfortunately, the solution of practical matrix equations often demands an exorbitant supply of computational resources, with many present-day simulation problems already stressing the limits of available computing systems.
For a matrix equation involving $N$ unknowns, the most sophisticated algorithms of the current day provide $\mathcal{O}\left(N\right)$ or $\mathcal{O}\left(N\text{polylog}N\right)$ runtime and memory complexity\cite{Chew,Chen,Omar},
and are heavily reliant on the structure of very specific problem classes.
Thus, as simulations naturally grow more complex with time, the requisite computational power must grow at least proportionately.

For such computationally intensive problems, quantum computers are growing in popularity as a potential remedy, 
as they are capable of leveraging procedures of dramatically reduced complexity relative to their classical counterparts\cite{Grover,Gerjuoy,Childs10}.
This is chiefly due to the manner in which quantum computers store and operate on data.
Values of interest are stored as the amplitudes of possible states for a register of qubits, meaning that a 1-qubit system (using a 2-state qubit) can store two values using the amplitudes of the $\ket{0}$ and $\ket{1}$ states; 
a 2-qubit system can store four values using the $\ket{00}$, $\ket{01}$, $\ket{10}$, and $\ket{11}$ states; and so on.
In general, the storage capacity of a quantum computer scales exponentially with the size of the system.
Quantum operators then act directly on the system's qubits, allowing the effects of operators to also scale exponentially with system size.

On the topic of matrix equation solvers, the Harrow/Hassidim/Lloyd (HHL) algorithm \cite{HHL} has become well established as the leading quantum algorithm.
For a system of condition number $\kappa$, this algorithm is capable of providing computational complexity $\mathcal{O}\left(\kappa^2\log\left(N\right)/\epsilon\right)$ in the calculation of a solution of precision $\epsilon$.
Thus, when $\kappa$ and $1/\epsilon$ scale logarithmically with $N$,
the procedure offers an exponential speedup over classical algorithms.

A detailed description of the HHL algorithm is given in section \ref{sec:HHL}, but we provide a brief overview of the procedure here for the sake of motivation.
In what follows, note that the notation $\ket{v}$ indicates a vector $\bar{v}$ with its components encoded onto the basis states of a quantum system as $\ket{v}=\sum_{j=0}^{N-1}v_i\ket{j}$, where $\bar{v}$ has $N$ components.
We typically suppress normalization factors where their meaning is irrelevant.
For simplicity, $\ket{v}$ is understood to also refer to the vector $\bar{v}$ itself.
We begin with a matrix equation $A\ket{x}=\ket{b}$, where $A$ is an $N\times N$ matrix, and $\ket{x}$ and $\ket{b}$ are $N$-dimensional vectors.
If a quantum system is initialized to the state $\ket{b}$, then its state will be a superposition of the eigenvectors of $A$:
\begin{equation}
    \ket{b}=\sum_{j=0}^{N-1} \beta_j\ket{u_j},
\end{equation}
where $\ket{u_j}$ is an eigenvector of $A$, and $\beta_j$ is the component of $\ket{b}$ along $\ket{u_j}$.
The solution vector $\ket{x}$ can then be obtained by applying the inverse of each eigenvalue $\lambda_j$ of $A$ to its corresponding eigenvector in the superposition $\ket{b}$:
\begin{equation}
    \sum_{j=0}^{N-1} \frac{\beta_j}{\lambda_j}\ket{u_j}=A^{-1}\ket{b}=\ket{x}.
\end{equation}
This application of inverse eigenvalues, conditioned on the presence of the corresponding eigenvectors, is the basic function of HHL.
By using quantum phase estimation (QPE) \cite{CleveQPE}, it is possible to extract and apply these eigenvalue factors in an efficient manner.
However, this phase estimation requires the implementation of a problem-dependent unitary in terms of basic quantum gates.
Generally the unitary $e^{iA}$ is employed for this purpose, but the process of decomposing this unitary into a sequence of basic quantum gates is nontrivial.
Therefore the process of implementing this operation on real quantum hardware presents a bottleneck for the procedure.
In this work we consider a particular method, inspired by research into quantum walks, which produces a unitary with a known decomposition into basic gates.

Quantum walks are a popular area of study in quantum computing, as they have shown success in aiding the construction of efficient algorithms \cite{Venegas-Andraca}.
Reference \cite{Kempe} gives an excellent introduction to the concept of quantum walks.
In essence, they are stochastic processes that, when evolved and measured, give information about the system in which they have evolved.
The structure of the evolution and the properties measured can be adjusted to change the information obtained.
We are interested in the walk procedure developed in \cite{Ambainis}, which considers the problem of element-distinctness.
This problem is quite far removed from our current study, but it is important for the fact that it introduces a particular walk operator based on Grover diffusion operators.
Following this research, \cite{Szegedy} further investigated diffusion-based quantum walks defined by probabilistic maps, and considered the spectra of the walk operators.
Finally, \cite{Childs10} expanded on these findings from the perspective of Hamiltonian simulation, and showed that the walk operators could be defined from a Hamiltonian to produce an operator with a spectrum closely 
related to that of the Hamiltonian itself.
Reference \cite{Childs12} further explored methods by which these walk operators could be implemented in practice.

The method developed in \cite{Childs10} uses QPE to extract the eigenvalues of the walk operator, which can then be related to the eigenvalues of the Hamiltonian.
This is the critical component from which we can develop a general procedure for solving matrix equations.
By treating a Hermitian system matrix as the Hamiltonian, we can apply QPE using the walk operator to extract the eigenvalues of the system matrix.
From there, the HHL algorithm can proceed as usual.

An algorithm presented in \cite{Kerenidis} also makes use of the relationship between quantum walk operators and the eigenvalues
of the system matrix (in particular, this work considers singular values) to produce an iterative quantum matrix solver.
Our procedure differs in that it provides a direct solution method, is more easily generalizable, and does not require explicit computation of any singular values of the system matrix.

The matrix solution procedure we present here is particularly efficient for cases involving sparse matrices, wherein it can easily be seen to outperform classical solvers.
This is due to the $\mathcal{O}\left(N_{nz}\log\left(N\right)\right)$ gate complexity of the associated solution circuit, where $N_{nz}$ is the total number of nonzero elements in the system matrix.
When system matrices exhibit little reliable structure, it is often impossible for classical procedures to provide performance better than $\mathcal{O}\left(N^3\right)$, regardless of the matrix sparsities.
A key example of this is the technique of sparse approximate inverse (SPAI) preconditioning \cite{Pan,Takahashi,Casacuberta}.
This technique typically generates a preconditioning matrix with $\mathcal{O}\left(N\right)$ nonzero elements, with a sparsity pattern that can be quite arbitrary, particularly for systems describing complex geometries.
This lack of matrix structure makes the development of efficient, general classical solvers extremely difficult, and often impossible.
However, our quantum method will always be capable of providing an inversion circuit having $\mathcal{O}\left(N\log\left(N\right)\right)$ complexity (so long as the matrix is well-conditioned).

Using the Qiskit SDK \cite{Qiskit}, we have developed a program \cite{Program} which closely follows the herein described procedure.
In the interest of expositional clarity, as well as to aid those seeking to investigate the program itself, we at times make reference to the specific functions of this program.
At these points, the program is referred to simply as ``the program''.

\section{The HHL Algorithm}\label{sec:HHL}
Here we describe the basic HHL algorithm which serves as a foundation for our matrix solution procedure.
For brevity, we study only a primitive form of the HHL algorithm.
More sophisticated formulations can be used to improve performance by, for example, ignoring ill-conditioned portions of the system matrix.

\subsection{Initialization}
Given a matrix $A\in\mathbb{C}^{N\times N}$ and a vector $\ket{b}\in\mathbb{C}^{N}$, the HHL algorithm is a quantum procedure to determine $\ket{x}$ such that
\begin{equation}\label{eq:HHL_init}
    A\ket{x}=\ket{b}.
\end{equation}
Due to the nature of quantum computation, several restrictive properties are required of this system.
It is important to note, however, that all of these properties can be satisfied for any input system by means of a simple preparation procedure.
Section \ref{sec:Arbitrary} considers such preparation in depth.
The restrictions are as follows:
The vector $\ket{b}$ is to be encoded onto a quantum system, and as such we require that $N$ be a power of two and that $\ket{b}$ be normalized.
Additionally, we require that $A$ is Hermitian.
This ensures that its eigenvectors form a basis for $\mathbb{C}^{N}$\cite{Horn-Johnson}, and also aids us in the definition of the QPE unitary.

The first step of the HHL procedure is the preparation of three registers:
\begin{equation}
    \ket{b}\ket{0}^{\otimes n_p}\ket{0}.
\end{equation}
The first register, which we refer to as the vector register, is initialized to $\ket{b}$, and hence it requires $n=\log_2N$ qubits.
The second register is an $n_p$-qubit phase register that is used to store the phase estimates produced by QPE.
The value of $n_p$ is chosen by the user, and it should be large enough to admit sufficient resolution in the phase estimation.
The final register is termed the ancilla register, and it consists of a single qubit.
Once the eigenvalues of $A$ have been estimated, the state of this register can be rotated to effect the application of the inverse eigenvalue factors.

The initial system state can be expressed in terms of the eigenvectors of $A$ as
\begin{equation}
    \ket{b}\ket{0}^{\otimes n_p}\ket{0} = 
    \sum_{j=0}^{N-1} \beta_j \ket{u_j}\ket{0}^{\otimes n_p}\ket{0},
\end{equation}
where $\ket{u_j}$ is the $j$-th eigenvector of $A$, and $\beta_j$ is the complex amplitude of $\ket{b}$ along $\ket{u_j}$.
The next step of the procedure is to apply QPE to this superposition of eigenvectors to determine the corresponding eigenphases.

\subsection{Quantum Phase Estimation}
\begin{figure}
\begin{center}
    \includegraphics[width=\columnwidth]{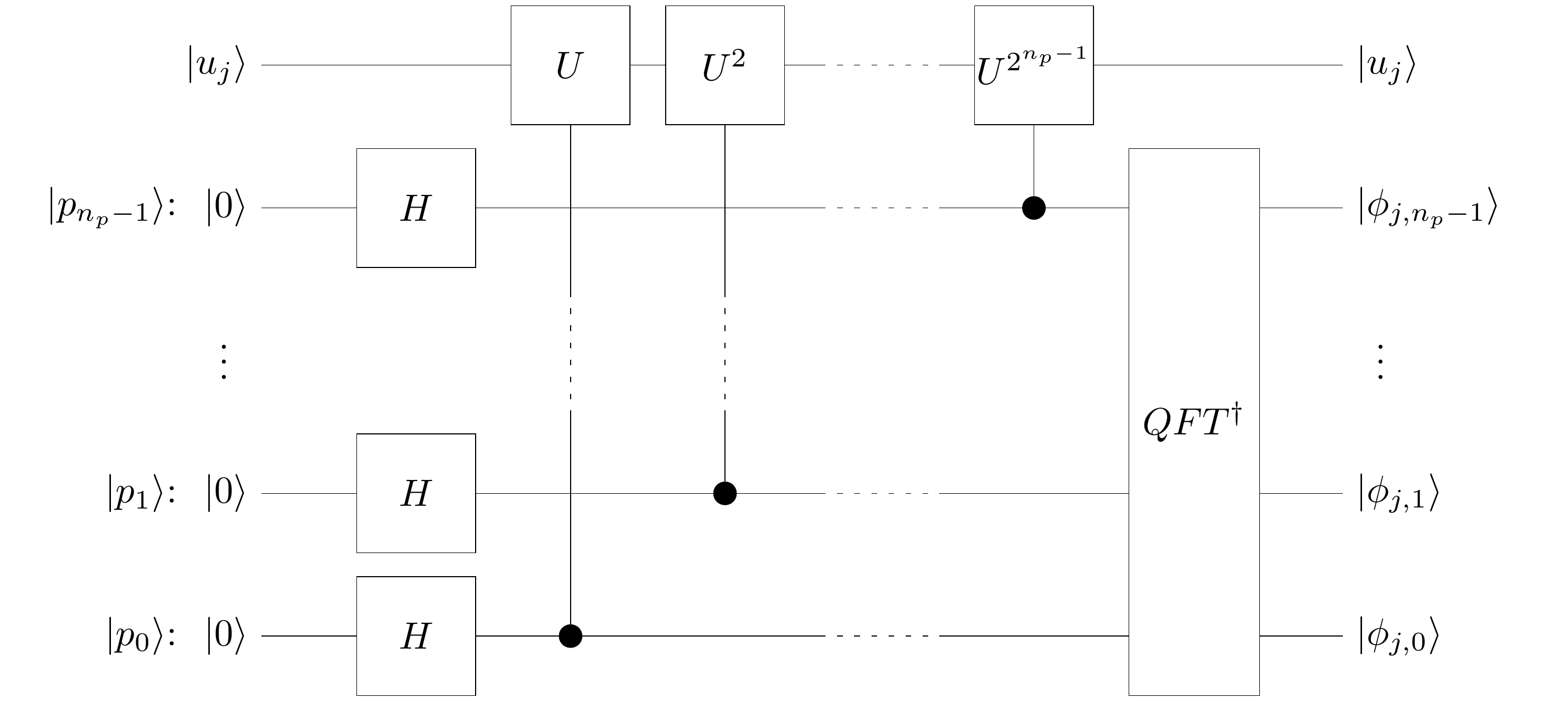}
    \caption{Diagram of quantum phase estimation.
             To emphasize the action of the eigenphase being read onto the phase register $\ket{p}$, the vector register is shown to have the state of a single eigenvector.
             In general, it will be prepared in a superposition of all eigenvectors of $U$.
             In the state $\ket{\phi_{j,a}}$, the second subscript $a$ indicates the the $a$-th bit of the eigenphase $\phi_j$ corresponding to $\ket{u_j}$.
             This final state assumes $\phi_j$ can be represented exactly by $n_p$ bits.
             When $\phi_j$ cannot be represented thusly, the final state of the phase register will have probabilities which spike sharply in the vicinity of the most accurate approximation to $\phi_j$.
             }
    \label{fig:QPE-basic}
\end{center}
\end{figure}
\begin{figure}
\begin{center}
    \includegraphics[width=\columnwidth]{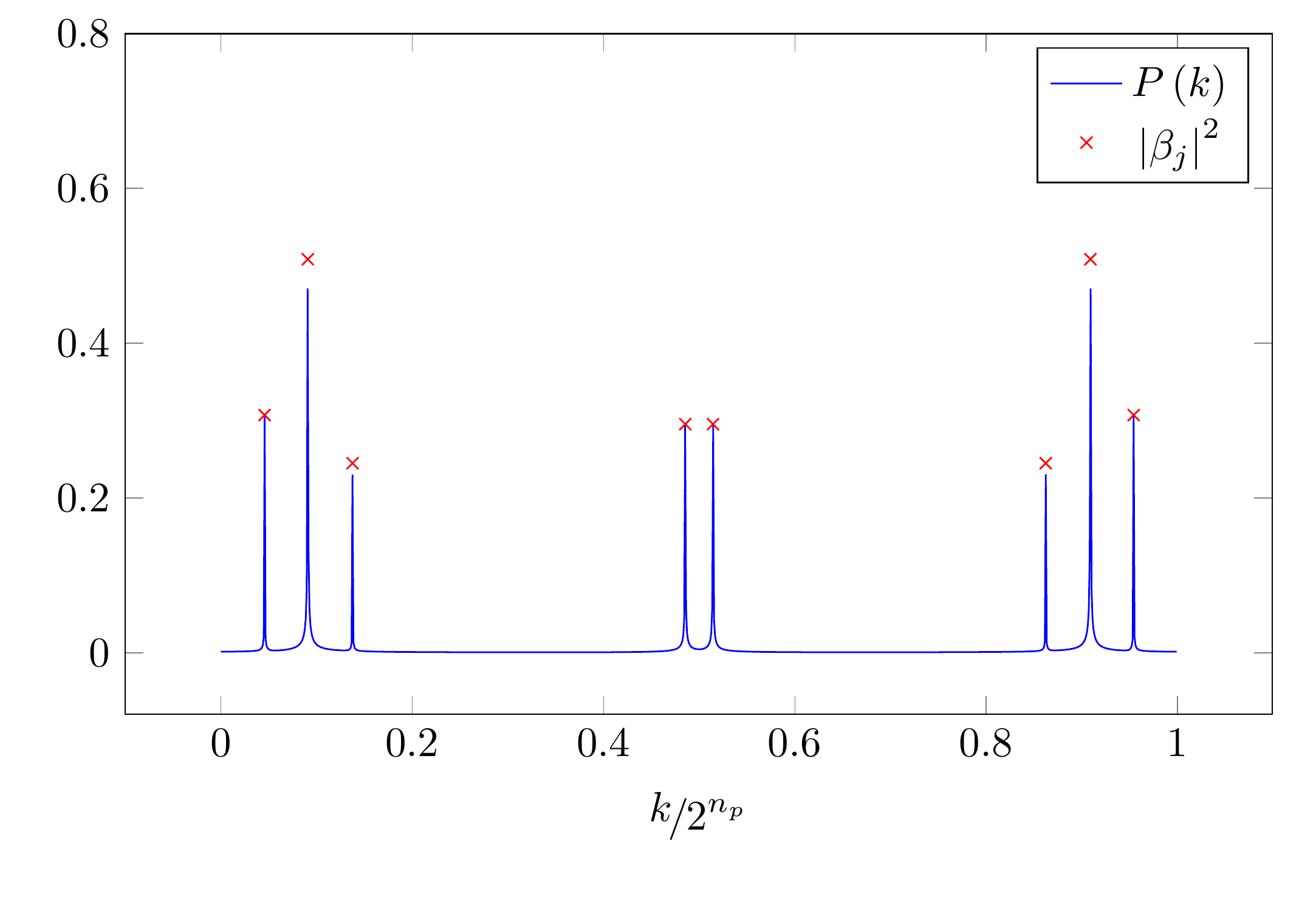}
    \caption{Illustration of the proportional presence of eigenphase estimations on the phase register after QPE.
             The unitary used for this calculation was the complex exponential $e^{iH}$ of a randomly generated Hermitian matrix $H$, and the operand vector was also randomly generated.
             $\left|\beta_j\right|^2$ gives the theoretical probability associated with eigenvector $\ket{u_j}$, and $P\left(k\right)$ gives the measured probability of observing the phase register in state $\ket{k}$.
             The amplitudes of phase states corresponding to accurate eigenphase approximations spike sharply and roughly in proportion to the presence of the eigenvectors,
             indicating that the state of a given phase estimate is entangled with the corresponding eigenstate.
             These spikes are not exactly proportional due to the probability being distributed over a nontrivial range of potential approximations.
             }
    \label{fig:Probs}
\end{center}
\end{figure}
Quantum phase estimation is a procedure for the estimation of the eigenphases of a unitary.
It begins with the observation that the eigenvalues of unitary operators lie on the complex unit circle, and hence they can be expressed in the form
\begin{equation}
    \mu_j=e^{2\pi i\phi_j},
\end{equation}
where $\mu_j$ is an eigenvalue of some unitary $U$, and $\phi_j\in\left[0,1\right)$ is termed the corresponding eigenphase.
The fundamental action of QPE is to use one register, initialized to an eigenvector $\ket{u_j}$ of $U$, to write a binary representation of the corresponding eigenphase onto an external phase register:
\begin{equation}
    \ket{u_j}\ket{0}^{\otimes n_p}
    \overset{\text{\tiny QPE}}{\longmapsto}
    \ket{u_j}\ket{\tilde{\phi}_j}.
\end{equation}
Here $\tilde{\phi}_j$ represents an $n_p$-bit approximation of $\phi_j$:
If $\phi_j$ can be exactly represented by $n_p$ bits, then the phase register will have the final state $\ket{\phi_j}$ exactly\footnote{Note that $\phi_j$ is not, in general, an integer.  The notation $\ket{\phi_j}$ is used here simply as a label to indicate a precise eigenphase estimation.  That is, $\ket{\phi_j}$ is the state $\ket{k}$ such that $k/2^{n_p}=\phi_j$. This notation follows naturally from (\ref{eq:phiapprox}).}.
Otherwise, the state of the phase register will have maximal amplitude at the best $n_p$-bit approximation of $\phi_j$, with the amplitudes of its other basis states decaying rapidly in the vicinity of this best approximation.
That is,
\begin{equation}\label{eq:phiapprox}
    \ket{\tilde{\phi}_j} = \sum_{k=1}^{2^{n_p}-1} \rho_{jk}\ket{k}.
\end{equation}
Where $\rho_{jk}$ spikes sharply in the vicinity of states with $k/2^{n_p}\approx \phi_j$.
We do not explore the specifics of the QPE procedure here, but a summary diagram is given in Fig. \ref{fig:QPE-basic}.

Since QPE is itself a linear operator, its effect when applied to a superposition of eigenvectors of $U$ is as follows:
\begin{equation}
    \sum_{j=0}^{N-1} \beta_j\ket{u_j}\ket{0}^{\otimes n_p}
    \overset{\text{\tiny QPE}}{\longmapsto}
    \sum_{j=0}^{N-1} \beta_j\ket{u_j}\ket{\tilde{\phi}_j}.
\end{equation}
That is, the approximated eigenphase states appear on the phase register entangled with their corresponding eigenvectors.
Fig. \ref{fig:Probs} provides an illustration of this behavior.

In the case of our current system, we require a unitary with eigenvalues and eigenvectors which can be easily related to those of $A$.
For the purposes of the HHL algorithm, the QPE unitary is typically defined as the exponential of the system matrix \cite{Hall},
\begin{equation}\label{eq:canonical}
    U = e^{iAt},
\end{equation}
where $t$ is some constant.
This operator is useful due to the close relationships between its eigenvalues and eigenvectors, and those of $A$.
However, it is interesting to note that this operator also corresponds to the simulation of a physical system subjected to a process described by the Hamiltonian $A$.
The evolution of such a system is given by Schr\"odinger's equation:
\begin{equation}
    i\hbar\frac{\partial\ket{\psi}}{\partial t} = A\ket{\psi}.
\end{equation}
When evolved from an initial state $\ket{\psi\left(0\right)}$ for a time $t$, the system's final state is given by
\begin{equation}
    \ket{\psi\left(t\right)} = e^{-iAt/\hbar}\ket{\psi\left(0\right)}.
\end{equation}
Thus, the operator in (\ref{eq:canonical}) can be considered to be a time-evolution operator for the Hamiltonian $A$, up to the factor of $-1/\hbar$.

Different values of $t$, and even combinations of different values, can be used to improve the performance of the calculation, but from this point on we will use $t=1$ for notational convenience.
Then the unitary $U=e^{iA}$ has eigenvalues $e^{i\lambda_j}$, and hence the eigenvalues of $A$ are related to the eigenphases of $U$ by
\begin{align}
    e^{i\lambda_j} = e^{2\pi i \phi_j} \implies 
    \lambda_j = 2\pi\phi_j + 2\pi l,
\end{align}
for some integer $l$.
Recall that $\phi_j$ is on the range $\left[0,1\right)$.
If we assume that $A$ has eigenvalues on the range $\left[-\pi,\pi\right]$ (this can be achieved in general by a simple rescaling of the system matrix) then we see that 
$\phi_j\in\left[0,\frac{1}{2}\right]$ correspond to $\lambda_j\in\left[0,\pi\right]$, and
$\phi_j\in\left(\frac{1}{2},1\right)$ correspond to $\lambda_j\in\left(-\pi,0\right)$.
Thus, $\lambda_j$ can be computed as
\begin{equation}
    \lambda_j = \begin{cases} 2\pi\phi_j, & \phi_j\in\left[0,\frac{1}{2}\right] \\
                              2\pi\left(\phi_j - 1\right), & \phi_j\in\left(\frac{1}{2},1\right). \end{cases}
\end{equation}
Of course, we do not compute the eigenphases exactly.
Instead, we transform the phase register to some superposition where the amplitude of a basis state $\ket{k}$ grows large when $k/N_p\approx\phi_j$.
Here we have defined $N_p=2^{n_p}$.
Then we in fact need to calculate potential approximations of the eigenvalues as
\begin{equation}\label{eq:HHL_lamapprox}
    \tilde{\lambda}_k = \begin{cases}
                            2\pi\frac{k}{N_p}, & k\le \frac{N_p}{2} \\
                            2\pi\left(\frac{k}{N_p}-1\right), & k> \frac{N_p}{2}.
                        \end{cases}
\end{equation}

From the above descriptions, we see that the effect of QPE on the current system state is
\begin{equation}
    \sum_{j=0}^{N-1} \beta_j \ket{u_j}\ket{0}^{\otimes n_p}\ket{0}
    \overset{\text{\tiny QPE}}{\longmapsto}
    \sum_{j=0}^{N-1}\sum_{k=0}^{N_p-1} \beta_j \rho_{jk} \ket{u_j}\ket{k}\ket{0},
\end{equation}
where $\rho_{jk}$ is the amplitude of the phase state $\ket{k}$ resulting from the action of QPE on eigenvector $\ket{u_j}$.
Thus, $\rho_{jk}$ should spike sharply when $\ket{k}$ nears an accurate approximation of the $j$-th eigenphase of $U$, and remain near zero otherwise.
With the eigenvalue extraction procedure defined, we can apply the third step of the procedure: ancilla rotation.

\subsection{Ancilla Rotation and Uncomputation}
In this step, the ancilla register is used to impose the inverse eigenvalue factors on the system.
We use the rotation operator
\begin{equation}
    R_y\left(2\arccos\left(\frac{C}{\tilde{\lambda}_k}\right)\right)\ket{0}
    = \frac{C}{\tilde{\lambda}_k}\ket{0} + \sqrt{1-\frac{C^{2}}{\tilde{\lambda}_k^2}}\ket{1},
\end{equation}
where $C$ is a constant chosen to ensure that all arguments of the $\arccos$ function obey its domain restrictions.
We recommend $C=2\pi/N_p$, as this corresponds to the minimum possible magnitude of the denominator $\tilde{\lambda}_k$ in (\ref{eq:HHL_lamapprox}), excepting $k=0$.
The $k=0$ case should be explicitly omitted from the calculation, as a nontrivial contribution from this state would indicate an eigenvalue near zero, and therefore an ill-conditioned matrix.
By using a controlled version of this rotation operator, with the phase register acting as the control register, 
it is possible to entangle the $\tilde{\lambda}_k$ rotation with the $\ket{k}$ state of the phase register.
Applying this controlled rotation to the ancilla register for all $k\in\left\{1,\dots,N_p-1\right\}$, the system is transformed to the state
\begin{equation}\label{eq:HHL_sysfinapprox}
    \sum_{j=0}^{N-1}\sum_{k=1}^{N_p-1} \beta_j\rho_{jk} \ket{u_j}\ket{k} \left(\frac{C}{\tilde{\lambda}_k}\ket{0} + \sqrt{1-\frac{C^{2}}{\tilde{\lambda}_k^2}}\ket{1}\right),
\end{equation}
where we have omitted the $k=0$ case on the assumption that $\rho_{j0}\approx 0$.

Note that the state (\ref{eq:HHL_sysfinapprox}) is approximately the following ideal state:
\begin{equation}\label{eq:HHL_sysfinexact}
    \sum_{j=0}^{N-1}\sum_{k=0}^{N_p-1} \beta_j\rho_{jk} \ket{u_j}\ket{k} \left(\frac{C}{\lambda_j}\ket{0} + \sqrt{1-\frac{C^{2}}{\lambda_j^2}}\ket{1}\right).
\end{equation}
This is true since $\rho_{jk}$ spikes sharply when $\tilde{\lambda}_k$ approaches $\lambda_j$, with the spike narrowing as more accurate approximations become available.
Then terms with $\tilde{\lambda}_k$ not approximately equal to $\lambda_j$ are nullified by their corresponding coefficients $\rho_{jk}$ approaching zero, 
and we can, up to an approximation error, replace $\tilde{\lambda}_k$ with $\lambda_j$, as we have done in (\ref{eq:HHL_sysfinexact}).
With the knowledge that the QPE error is obscured by this step, we from this point on assume that the system is in state (\ref{eq:HHL_sysfinexact}) exactly.
Acting on this state with an inverse QPE will then ``uncompute'' the phase register, leaving us in the state
\begin{equation}\label{eq:HHL_res}
    \sum_{j=0}^{N-1} \beta_j \ket{u_j}\ket{0}^{\otimes n_p} \left(\frac{C}{\lambda_j}\ket{0} + \sqrt{1-\frac{C^{2}}{\lambda_j^2}}\ket{1}\right).
\end{equation}
This is the final state of the system, from which our solution can be extracted.
Selecting those states with $\ket{0}$ in all qubits of the phase register and the ancilla register, we have
\begin{equation}\label{eq:HHL_fin}
    C\sum_{j=0}^{N-1} \frac{\beta_j}{\lambda_j} \ket{u_j}.
\end{equation}
This state, when divided by $C$, is exactly the decomposition of $A^{-1}\ket{b}$ in terms of the eigenvectors of $A$, and hence (\ref{eq:HHL_fin}) gives our final solution $\ket{x}$.

\section{The Walk Operator}\label{sec:WalkOp}
A walk operator is a unitary operator that is used during a quantum walk procedure, with each application of the operator representing a ``step'' made by the system.
As mentioned above, we do not here describe a quantum walk procedure per se.
Rather, we simply adopt a walk operator from a formerly studied quantum walk procedure \cite{Ambainis,Szegedy}.
The operator chosen is useful because its implementation is well defined for any arbitrary system matrix.

The walk operator itself has several constituents which must be defined before the operator can be stated succinctly.
To begin, we define the vectors
\begin{equation}
    \ket{\phi_j} = \frac{1}{\sqrt{N}}\sum_{k=0}^{N-1} \ket{k}\sqrt{\frac{N}{X}A_{jk}^{*}},
\end{equation}
for $j\in\left\{0,\dots,N-1\right\}$.
Here $X$ is a constant which must satisfy $X\ge N\left|A_{jk}\right|_{\text{max}}$, where $\left|A_{jk}\right|_{\text{max}}=\max_{j,k}\left(\left|A_{jk}\right|\right)$.
This factor is required for the purposes of ancilla rotation, as discussed in section \ref{sec:Practical}.
The $\ket{\phi_j}$ vectors are useful to us as carriers of information about the system matrix, and they form foundational elements of the walk operator.
However, they are possessed of some critical shortcomings.
First, they are not normalized.
While they could be explicitly normalized through appropriate $j$-dependent factors, subsequent derivations serve to show that this scheme would destroy crucial operator properties.
In particular, the result of (\ref{eq:TdagST-result}) would no longer hold.
The second concern regarding the $\ket{\phi_j}$ vectors is their preparation.
Producing unitary operators to generate these states is a nontrivial problem, and even once computed, these operators would also carry $j$-dependent normalization factors.

Both of these shortcomings can be remedied with relative ease by introducing an ancillary qubit and studying the set of augmented states
\begin{align}
    \ket{\phi_j^a} = \frac{1}{\sqrt{N}}\sum_{k=0}^{N-1} \ket{k}\Big[&\sqrt{\frac{N}{X}A_{jk}^{*}}\ket{0} 
    +\sqrt{1-\frac{N}{X}\left|A_{jk}\right|}\ket{1}\Big].
\end{align}
These states are normalized, and they can be easily produced through unitary transformations (see section \ref{sec:Practical}).
Note that the $\ket{\phi_j}$ states appear when the ancilla is in the $\ket{0}$ state.
Hence, the $\ket{1}$ state of the ancilla can be interpreted as a failure state.

Storage of each $\ket{\phi_j^a}$ state requires an $\left(n+1\right)$-qubit register.
We define two such registers, and refer to them by the names $\ket{r_1}$ and $\ket{r_2}$.
The ancilla-augmented basis states for these registers are indicated by a superscript $a$, e.g. $\ket{j^a}$.
Note that there are $2^{n+1}=2N$ augmented basis states.
When necessary, we always assume that the ancillary qubit is appended to the relevant register.
In this procedure, there are two basic operators which act on these registers.
The first is the swap operator, which swaps the content of $\ket{r_1}$ and $\ket{r_2}$:
\begin{equation}
    S = \sum_{j^a=0}^{2N-1}\sum_{k^a=0}^{2N-1} \ket{k^a}\ket{j^a}\bra{j^a}\bra{k^a}.
\end{equation}
The second operator is that which produces $\ket{\phi_j^a}$ on $\ket{r_2}$ when $\ket{r_1}$ is in the state $\ket{j}$ with ancilla state $\ket{0}$:
\begin{equation}
    T = \sum_{j=0}^{N-1} \left(\ket{j,0}\ket{\phi_j^a}\bra{j,0}+\ket{j,1}\ket{\zeta_j^a}\bra{j,1}\right).
\end{equation}
Here the $\ket{\zeta_j^a}$ states correspond to failure states.
A forthcoming analysis of the $T^{\dagger}ST$ operator shows that they must have ancilla state $\ket{1}$,
but otherwise their definition is insignificant.
These are proper quantum states, and hence they must be normalized.

These prerequisites are sufficient for us to state the walk operator:
\begin{equation}
    W = iS\left(2TT^{\dagger}-I\right).
\end{equation}
The process of implementing $W$ is further explored in section \ref{sec:Practical}.
For the remainder of this section, we take it as given that $W$ can be implemented.
As stated above, this walk operator is particularly interesting to us because it has eigenvalues and eigenvectors closely related to those of $A$ itself.
In deriving these relationships between $W$ and $A$, we first calculate some useful relationships for the constituent operators $T$ and $S$.
To begin, consider the product $T^{\dagger}T$:
\begin{align}
    T^{\dagger}T 
    &= \left[\sum_{j=0}^{N-1} \left(\ket{j,0}\bra{j,0}\bra{\phi_j^a} + \ket{j,1}\bra{j,1}\bra{\zeta_j^a}\right)\right]
    \nonumber \\
    &\hphantom{=} \cdot\left[\sum_{k=0}^{N-1} \left(\ket{k,0}\ket{\phi_k^a}\bra{k,0} + \ket{k,1}\ket{\zeta_k^a}\bra{k,1}\right)\right]
    \nonumber \\
    &= \sum_{j=0}^{N-1}\sum_{k=0}^{N-1} \ket{j,0}\braket{\phi_j^a|\phi_k^a}\braket{j,0|k,0}\bra{k,0}
    \nonumber \\
    &\hphantom{=\sum_{j=0}^{N-1}\sum_{k=0}^{N-1}} + \ket{j,1}\braket{\zeta_j^a|\zeta_k^a}\braket{j,1|k,1}\bra{k,1}.
\end{align}
Recalling that the $\ket{\phi_j^a}$ and $\ket{\zeta_j^a}$ states are normalized and that the basis states are orthonormal, we have
\begin{align}\label{eq:TdagT}
    T^{\dagger}T 
    = \sum_{j=0}^{N-1} \left(\ket{j,0}\bra{j,0} + \ket{j,1}\bra{j,1}\right)
    = I.
\end{align}
Note that despite the above, $TT^{\dagger}$ does not yield the identity operator, and hence $T$ is not a unitary transformation as currently stated.
The second relationship of interest is the following:
\begin{align}\label{eq:TdagST-begin}
    T^{\dagger}ST 
    &= \left[\sum_{j=0}^{N-1} \left(\ket{j,0}\bra{j,0}\bra{\phi_j^a} + \ket{j,1}\bra{j,1}\bra{\zeta_j^a}\right) \right]
    \nonumber \\
    &\hphantom{=} \cdot\left[\sum_{k=0}^{N-1} \left(\ket{\phi_k^a}\ket{k,0}\bra{k,0} + \ket{\zeta_k^a}\ket{k,1}\bra{k,1}\right) \right]
    \nonumber \\
    &= \sum_{j=0}^{N-1}\sum_{k=0}^{N-1} \Big(\ket{j,0}\braket{\phi_j^a|k,0}\braket{j,0|\phi_k^a}\bra{k,0}
    \nonumber \\
    &\hphantom{=\sum_{j=0}^{N-1}} +\ket{j,0}\braket{\phi_j^a|k,1}\braket{j,0|\zeta_k^a}\bra{k,1}
    \nonumber \\
    &\hphantom{=\sum_{j=0}^{N-1}} +\ket{j,1}\braket{\zeta_j^a|k,0}\braket{j,1|\phi_k^a}\bra{k,0}
    \nonumber \\
    &\hphantom{=\sum_{j=0}^{N-1}} +\ket{j,1}\braket{\zeta_j^a|k,1}\braket{j,1|\zeta_k^a}\bra{k,1}\Big).
\end{align}
If this operator acts on an arbitrary input vector $\ket{\psi,0}$, we have
\begin{align}
    &T^{\dagger}ST \ket{\psi,0}
    \nonumber \\
    &= \sum_{j=0}^{N-1}\sum_{k=0}^{N-1} \Big(\ket{j,0}\braket{\phi_j^a|k,0}\braket{j,0|\phi_k^a}\braket{k,0|\psi,0}
    \nonumber \\
    &\hphantom{=\sum_{j=0}^{N-1}\sum_{k=0}^{N-1}} +\ket{j,1}\braket{\zeta_j^a|k,0}\braket{j,1|\phi_k^a}\braket{k,0|\psi,0}\Big).
\end{align}
If each $\ket{\zeta_j^a}$ has ancilla state $\ket{1}$, this expression becomes
\begin{align}
    T^{\dagger}ST \ket{\psi,0}
    &= \sum_{j=0}^{N-1}\sum_{k=0}^{N-1} \ket{j,0}\braket{\phi_j^a|k,0}
    \nonumber \\
    &\hphantom{=====.} \cdot\braket{j,0|\phi_k^a}\braket{k,0|\psi,0}.
\end{align}
This elimination of the $\ket{\zeta_j^a}$ contributions is a crucial result, and hence we require that these states have ancilla state $\ket{1}$.
The inner products are
\begin{align}\label{eq:inner1}
    &\braket{\phi_j^a|k,0}
    \nonumber \\
    &= \Bigg[\frac{1}{\sqrt{N}}\sum_{p=0}^{N-1} \bra{p}\Big(\sqrt{\frac{N}{X}}\left(\sqrt{A_{jp}^*}\right)^*\bra{0}
    \nonumber \\
    &\hphantom{=\Big[\frac{1}{\sqrt{N}}\sum_{p=0}^{N-1}\Big(} +\sqrt{1-\frac{N}{X}\left|A_{jp}\right|}\bra{1}\Big)\Bigg] \ket{k}\ket{0}
    \nonumber \\
    &= \left(\sqrt{\frac{A_{jk}^*}{X}}\right)^*.
\end{align}
\begin{align}
    &\braket{j,0|\phi_k^a}
    \nonumber \\
    &= \bra{0}\bra{j}\Bigg[\frac{1}{\sqrt{N}}\sum_{p=0}^{N-1}\ket{p}\Big(\sqrt{\frac{N}{X}A_{kp}^{*}}\ket{0}
    \nonumber \\
    &\hphantom{=\bra{0}\bra{j}\Big[\frac{1}{\sqrt{N}}\sum_{p=0}^{N-1}\ket{p}\Big(} +\sqrt{1-\frac{N}{X}\left|A_{kp}\right|}\ket{1}\Big)\Bigg]
    \nonumber \\
    &= \sqrt{\frac{A_{kj}^{*}}{X}}.
\end{align}
Then we have
\begin{align}\label{eq:TdagST-result}
    &T^{\dagger}ST \ket{\psi,0}
    \nonumber \\
    &= \sum_{j=0}^{N-1}\sum_{k=0}^{N-1} \ket{j,0} \sqrt{\frac{A_{kj}^{*}}{X}} \left(\sqrt{\frac{A_{jk}^*}{X}}\right)^* \braket{k,0|\psi,0}.
    \nonumber \\
    &= \sum_{j=0}^{N-1}\sum_{k=0}^{N-1} \ket{j,0}\frac{A_{jk}}{X}\braket{k,0|\psi,0}.
    \nonumber \\
    &=\left(\frac{1}{X}A\ket{\psi}\right)\ket{0}.
\end{align}
The statement that $\sqrt{A_{kj}^{*}}\left(\sqrt{A_{jk}^*}\right)^* = A_{jk}$ is true, but it presents an issue for implementation when $A_{jk}$ lies on the negative real axis.
This problem is further discussed in section \ref{sec:Practical}.
By the above, we see that $T^{\dagger}ST$, when applied to a vector with ancilla state $\ket{0}$, acts as a rescaled version of $A$.
We introduce the shorthand 
\begin{equation}
    \hat{A}\equiv\frac{1}{X}\sum_{j=0}^{N-1}\sum_{k=0}^{N-1} A_{jk}\ket{j,0}\bra{k,0}.
\end{equation}
This transformation is a $2N\times2N$ matrix, and it can therefore be applied directly to the $2N$-dimensional augmented states.
This simplifies the statement of (\ref{eq:TdagST-result}):
\begin{equation}
    T^{\dagger}ST \ket{\psi,0} = \hat{A}\ket{\psi,0}.
\end{equation}

Now we can begin the analysis of the walk operator's eigenvalues and eigenvectors.
We begin with the assertion that the eigenvalues of $\hat{A}$ must lie on the range $\left[-1,1\right]$.
By the definition of $X$, this is in fact always satisfied regardless of the choice of $A$, as we show in section \ref{sec:Arbitrary}.
With this established, consider now the effect of applying $W$ to the vector $T\ket{u_j^a}=T\ket{u_j,0}$ (recall that $\ket{u_j}$ is an eigenvector of $A$):
\begin{equation}\label{eq:WT}
    WT\ket{u_j^a} = iST\ket{u_j^a}.
\end{equation}
Note that we have continued to assume that the input vector is provided with ancilla state $\ket{0}$.
Meanwhile, if $W$ is applied to $ST\ket{u_j^a}$, we obtain
\begin{equation}\label{eq:WST}
    WST\ket{u_j^a} = 2i\hat{\lambda}_jST\ket{u_j^a} - iT\ket{u_j^a},
\end{equation}
where $\hat{\lambda}_j=\lambda_j/X$ is the eigenvalue of $\hat{A}$ corresponding to $\ket{u_j^a}$.
Thus, if $W$ is applied to any superposition of $T\ket{u_j^a}$ and $ST\ket{u_j^a}$, the result will be a simple superposition of the same two vectors:
\begin{align}
    &W\left(T\ket{u_j^a} + \gamma ST\ket{u_j^a}\right)
    \nonumber \\
    &= -i\gamma T\ket{u_j^a} + i\left(1+2\hat{\lambda}_j\gamma\right)ST\ket{u_j^a},
\end{align}
where $\gamma$ is an arbitrary factor for the relative phase and magnitude of the two contributions.
Then $\left(T\ket{u_j^a} + \gamma ST\ket{u_j^a}\right)$ is an eigenvector $\ket{v_j}$ of $W$, with eigenvalue $\mu_j$, if $\gamma=i\mu_j$ and $i\mu_j^2=i\left(1+2i\hat{\lambda}_j\mu_j\right)$.
Solving this expression for the eigenvalues, we find two solutions:
\begin{equation}\label{eq:evals}
    \mu^{\pm}_j = i\hat{\lambda}_j\pm\sqrt{1-\hat{\lambda}_j^2}.
\end{equation}
The eigenvectors $\ket{v_j^{\pm}}$ corresponding to these eigenvalues can be normalized by computing the inner products of their un-normalized counterparts:
\begin{align}
    &\braket{v_j^{\pm}|v_j^{\pm}}
    \nonumber \\
    &= \left(\bra{u_j^a}T^{\dagger} - i\left(\mu_j^{\pm}\right)^{*} \bra{u_j^a}T^{\dagger}S^{\dagger}\right)\left(T\ket{u_j^a} + i\mu_j^{\pm} ST\ket{u_j^a}\right)
    \nonumber \\
    &= \bra{u_j^a}T^{\dagger}T\ket{u_j^a} + i\mu_j^{\pm}\bra{u_j^a}T^{\dagger}ST\ket{u_j^a} 
    \nonumber \\
    &\hphantom{=} - i\left(\mu_j^{\pm}\right)^{*}\bra{u_j^a}T^{\dagger}S^{\dagger}T\ket{u_j^a} + \left|\mu_j^{\pm}\right|^2\bra{u_j^a}T^{\dagger}S^{\dagger}ST\ket{u_j^a}
    \nonumber \\
    &= \braket{u_j^a|u_j^a} + i\mu_j^{\pm}\bra{u_j^a}\hat{A}\ket{u_j^a} 
    \nonumber \\
    &\hphantom{=} - i\left(\mu_j^{\pm}\right)^{*}\bra{u_j^a}\hat{A}^{\dagger}\ket{u_j^a} + \left|\mu_j^{\pm}\right|^2\braket{u_j^a|u_j^a}
    \nonumber \\
    &= 1 + i\mu_j^{\pm}\hat{\lambda}_j - i\left(\mu_j^{\pm}\right)^{*}\hat{\lambda}_j + 1
    \nonumber \\
    &= 2\left(1-\hat{\lambda}_j^2\right).
\end{align}
Therefore, $W$ has the normalized eigenvectors
\begin{equation}\label{eq:evecs}
    \ket{v_j^{\pm}} = \frac{1+i\mu^{\pm}_jS}{\sqrt{2\left(1-\hat{\lambda}_j^2\right)}} T\ket{u_j^a}.
\end{equation}

While the above analysis does not constitute a comprehensive investigation into the eigenvalues and eigenvectors of $W$, this simplified derivation is, as shown in what follows, sufficient for our purposes.
For a detailed analysis, see \cite{Szegedy}.
To demonstrate the sufficiency of the above analysis, consider the following superposition of eigenvectors of $W$:
\begin{align}
    \frac{\left(1+i\hat{\lambda}_j\mu_j^-\right)\ket{v_j^+} + \left(1+i\hat{\lambda}_j\mu_j^+\right)\ket{v_j^-}}{\sqrt{2\left(1-\hat{\lambda}_j^2\right)}}.
\end{align}
Expanding the eigenvector expressions and simplifying, we find
\begin{align}
    &\frac{1}{2\left(1-\hat{\lambda}_j^2\right)}
    \Big[\left(1+i\hat{\lambda}_j\mu_j^-\right)\left(1+i\mu_j^+S\right)T\ket{u_j^a}
    \nonumber \\
    &\hphantom{\frac{1}{2\left(1-\hat{\lambda}_j^2\right)}\Big[}+ \left(1+i\hat{\lambda}_j\mu_j^+\right)\left(1+i\mu_j^-S\right)T\ket{u_j^a}\Big]
    \nonumber \\
    &=\frac{1}{2\left(1-\hat{\lambda}_j^2\right)}
    \Big[2 + i\left(\mu_j^++\mu_j^-\right)S 
    \nonumber \\
    &\hphantom{=\frac{1}{2\left(1-\hat{\lambda}_j^2\right)}\Big[} + i\hat{\lambda}_j\left(\mu_j^++\mu_j^-\right) - 2\hat{\lambda}\mu_j^+\mu_j^-S\Big]T\ket{u_j^a}
    \nonumber \\
    &=\frac{1}{2\left(1-\hat{\lambda}_j^2\right)}
    \left[2 - 2\hat{\lambda}_jS - 2\hat{\lambda}_j^2+2\hat{\lambda}_jS\right]T\ket{u_j^a}
    \nonumber \\
    &=T\ket{u_j^a}.
\end{align}
This result gives us the necessary components to build a matrix equation solver.
By applying $T$ to the initial right-hand side vector $\ket{b,0}$, we can transform the vector into a predictable superposition of the $\ket{v_j^{\pm}}$'s:
\begin{align}\label{eq:evec-trans}
    T\ket{b,0}
    &= \sum_{j=0}^{N-1} \beta_jT\ket{u_j^a}
    \nonumber \\
    &= \sum_{j=0}^{N-1} \frac{\beta_j}{\sqrt{2\left(1-\hat{\lambda}_j^2\right)}} \Big[\left(1+i\hat{\lambda}_j\mu_j^-\right)\ket{v_j^+} 
    \nonumber \\
    &\hphantom{\sum_{j=0}^{N-1} \frac{\beta_j}{\sqrt{2\left(1-\hat{\lambda}_j^2\right)}}\Big[} + \left(1+i\hat{\lambda}_j\mu_j^+\right)\ket{v_j^-}\Big].
\end{align}
Then an application of QPE to this state using $W$ as the requisite unitary yields the phase corresponding to $\mu_j^+$ or $\mu_j^-$ when $\ket{u_j^a}$ appears in the initial state.
From here it is possible to extract $\lambda_j$ and apply an HHL-style ancilla rotation.
After applying inverse QPE, the final step of the procedure is to uncompute by applying $T^{\dagger}$, thereby leaving only the rotated initial state, as in the result of HHL.

\section{Considerations for Practical Implementation}\label{sec:Practical}
The analysis we have provided so far represents a complete theoretical formulation of the procedure, 
but practical implementation presents several nontrivial problems.
In particular, the implementation of $W$, the initial application of $T$, and the eigenvalue extraction require elaboration.
Here we present a recommended approach to the resolution of these sticking points.

Implementation of $W$ requires, in addition to the straightforward operators $i$ and $S$, the more involved operator $2TT^{\dagger}-I$.
This operator is a sum of conditional reflectors\footnote{A brief introduction to the concept of reflectors is given in appendix \ref{sec:Reflectors}.} about the $\ket{\phi_j^a}$ and $\ket{\zeta_j^a}$ states, as the following calculation shows.
We begin by expanding and simplifying the $TT^{\dagger}$ portion of the operator:
\begin{align}
    TT^{\dagger}&=\left[\sum_{j=0}^{N-1} \left(\ket{j,0}\ket{\phi_j^a}\bra{j,0}+\ket{j,1}\ket{\zeta_j^a}\bra{j,1}\right)\right]
    \nonumber \\
    &\hphantom{=}\cdot \left[\sum_{k=0}^{N-1} \left(\ket{k,0}\bra{k,0}\bra{\phi_k^a}+\ket{k,1}\bra{k,1}\bra{\zeta_k^a}\right)\right]
    \nonumber \\
    &= \sum_{j=0}^{N-1}\sum_{k=0}^{N-1} \Big[\braket{j,0|k,0}\left(\ket{j,0}\bra{k,0}\otimes\ket{\phi_j^a}\bra{\phi_k^a}\right)
    \nonumber \\
    &\hphantom{= \sum_{j=0}^{N-1}\sum_{k=0}^{N-1}\Big[} +\braket{j,1|k,1}\left(\ket{j,1}\bra{k,1}\otimes\ket{\zeta_j^a}\bra{\zeta_k^a}\right)\Big]
    \nonumber \\
    &= \sum_{j=0}^{N-1} (\ket{j,0}\bra{j,0}\otimes\ket{\phi_j^a}\bra{\phi_j^a} 
    \nonumber \\
    &\hphantom{=\sum_{j=0}^{N-1} (} + \ket{j,1}\bra{j,1}\otimes\ket{\zeta_j^a}\bra{\zeta_j^a}).
\end{align}
Then the full operator $2TT^{\dagger}-I$ is
\begin{align}
    &2\sum_{j=0}^{N-1} \left(\ket{j,0}\bra{j,0}\otimes\ket{\phi_j^a}\bra{\phi_j^a} + \ket{j,1}\bra{j,1}\otimes\ket{\zeta_j^a}\bra{\zeta_j^a}\right)
    \nonumber \\
    &- \sum_{j=0}^{N-1} \left(\ket{j,0}\bra{j,0}\otimes I + \ket{j,1}\bra{j,1}\otimes I\right)
    \nonumber \\
    &= \sum_{j=0}^{N-1} \Big[\ket{j,0}\bra{j,0}\otimes\left(2\ket{\phi_j^a}\bra{\phi_j^a}-I\right)
    \nonumber \\
    &\hphantom{=\sum_{j=0}^{N-1} \Big[} + \ket{j,1}\bra{j,1}\otimes\left(2\ket{\zeta_j^a}\bra{\zeta_j^a}-I\right)\Big].
\end{align}
That is, the operator reflects $\ket{r_2}$ about $\ket{\phi_j^a}$ when $\ket{r_1}$ is in the state $\ket{j,0}$, or it reflects about $\ket{\zeta_j^a}$ when $\ket{r_1}$ is in the state $\ket{j,1}$.
To implement this operator, we first consider a state preparation operator $B_j$ which prepares $\ket{\phi_j^a}$ from the $\ket{0}$ state.
By applying, in sequence, the inverse of this state preparation operator, a reflection about the $\ket{0}$ state, and finally the state preparation operator, we find
\begin{align}
    &B_j\left(2\ket{0}\bra{0}-I\right)B_j^{\dagger}
    \nonumber \\
    &= 2B_j\ket{0}\bra{0}B_j^{\dagger}-B_jB_j^{\dagger}
    \nonumber \\
    &= 2\ket{\phi_j^a}\bra{\phi_j^a}-I.
\end{align}
Thus, this procedure effectively reflects about the state $\ket{\phi_j^a}$.
Likewise, we can reflect about the $\ket{\zeta_j^a}$ states using an operator $B_j^{\prime}$ which prepares $\ket{\zeta_j^a}$ from $\ket{0}$:
\begin{equation}
    B_j^{\prime}\left(2\ket{0}\bra{0}-I\right)\left(B_j^{\prime}\right)^{\dagger}
    =
    2\ket{\zeta_j^a}\bra{\zeta_j^a}-I.
\end{equation}
The full operator can then be implemented as
\begin{align}
    &2TT^{\dagger}-I
    \nonumber \\
    &= \sum_{j=0}^{N-1} \Big[\ket{j,0}\bra{j,0}\otimes\left(B_j\left(2\ket{0}\bra{0}-I\right)B_j^{\dagger}\right)
    \nonumber \\
    &\hphantom{=\sum_{j=0}^{N-1}} + \ket{j,1}\bra{j,1}\otimes\left(B_j^{\prime}\left(2\ket{0}\bra{0}-I\right)\left(B_j^{\prime}\right)^{\dagger}\right)\Big].
\end{align}
Note that the swap operations involved in the walk operator mean that, at intermediate stages, the system stores essential data on the $\ket{1}$ state of the ancilla,
and hence the reflections about the $\ket{\zeta_j}$'s cannot be ignored.

The implementation of the state preparation operators is straightforward.
For the $B_j$ operators, we begin by preparing a uniform superposition:
\begin{equation}
    \frac{1}{\sqrt{N}} \sum_{k=0}^{N-1} \ket{k,0}.
\end{equation}
We can then apply a single Pauli $X$ gate to the ancilla bit to obtain
\begin{equation}
    \frac{1}{\sqrt{N}} \sum_{k=0}^{N-1} \ket{k,1}.
\end{equation}
Note that, for $A_{jk}=0$, the amplitudes of the $\ket{k^a}$ states are now exactly as desired.
Then for $\ket{k}$ corresponding to nonzero $A_{jk}$, we rotate the ancilla to obtain:
\begin{equation}\label{eq:PreppedState}
    \frac{1}{\sqrt{N}} \sum_{k=0}^{N-1} \ket{k} \left[\sqrt{\frac{N}{X}A_{jk}^*}\ket{0} + \sqrt{1-\frac{N}{X}\left|A_{jk}\right|}\ket{1}\right].
\end{equation}
It is at this point that the role of $X$ becomes clear.
In order to ensure a valid rotation, we select $X$ such that $X/N\ge\left|A_{jk}\right|_{\text{max}}$.
Note that the calculation of $\sqrt{A_{jk}^*}$ requires some additional consideration.
In particular, the result of (\ref{eq:TdagST-result}) requires that
\begin{equation}
    \sqrt{A_{kj}^{*}}\left(\sqrt{A_{jk}^*}\right)^* = A_{jk}.
\end{equation}
In most cases, this can be satisfied by taking
\begin{equation}\label{eq:root-def}
    \sqrt{A^*_{jk}} = \sqrt{\left|A_{jk}\right|}e^{-i\arg\left(A_{jk}\right)/2},
\end{equation}
and using the principal square root for $\sqrt{\left|A_{jk}\right|}$.
However, if $A_{jk}\in\left(-\infty,0\right)$, then $A_{jk}=A_{kj}$, and we have
\begin{align}
    \sqrt{A_{kj}^{*}}\left(\sqrt{A_{jk}^*}\right)^* &= \left(\sqrt{\left|A_{jk}\right|}e^{-i\pi/2}\right)\left(\sqrt{\left|A_{jk}\right|}e^{i\pi/2}\right) 
    \nonumber \\
    &= \left|A_{jk}\right|.
\end{align}
For $j\ne k$, we can recover the correct sign by taking the negative square root in (\ref{eq:root-def}) whenever $k>j$.
That is, if $A_{jk}$ is a negative real number, we use
\begin{equation}
    \sqrt{A^*_{jk}} = \text{sign}\left(j-k\right)\sqrt{\left|A_{jk}\right|}e^{-i\arg\left(A_{jk}\right)/2}.
\end{equation}
While this prescription is effective for $j\ne k$, it does not suffice when $j=k$.
To address this case, we simply prevent negative elements from appearing on the diagonal of $A$ by adding an appropriate multiple of the identity matrix.
Proper treatment of this modification is considered in section \ref{sec:Arbitrary}.

The $B_j^{\prime}$ operators can be implemented by simply switching the ancilla state of the operand register to $\ket{1}$.
This gives $\ket{\zeta_j^a}=\ket{0}^{\otimes n}\ket{1}$, which is sufficient for our purposes.

The initial application of $T$ is also of concern.
The operation is defined such that, when the $\ket{r_1}$ ancilla qubit is in the $\ket{0}$ state, the $\ket{\phi_j^a}$ states are produced on $\ket{r_2}$ regardless of the initial state of the register.
This is problematic to implement in practice, but since the only direct application of $T$ occurs at the beginning of the calculation, we can safely assume that $\ket{r_2}$ begins with all qubits in the $\ket{0}$ state.
Then the initial application of $T$, which we shall distinguish by the term $T_0$, can be effectively applied by conditional application of the state preparation operators:
\begin{align}
    T_0 = \sum_{j=0}^{N-1} \left(\ket{j,0}\bra{j,0}\otimes B_j + \ket{j,1}\bra{j,1}\otimes B_j^{\prime}\right).
\end{align}
Unlike $T$, this operator is unitary, and hence it is suitable for use in a quantum circuit.
Since the initial state must be supplied with ancilla state $\ket{0}$, it is also possible to ignore the applications of $B_j^{\prime}$.

\begin{figure}
\begin{center}
    \includegraphics[width=\columnwidth]{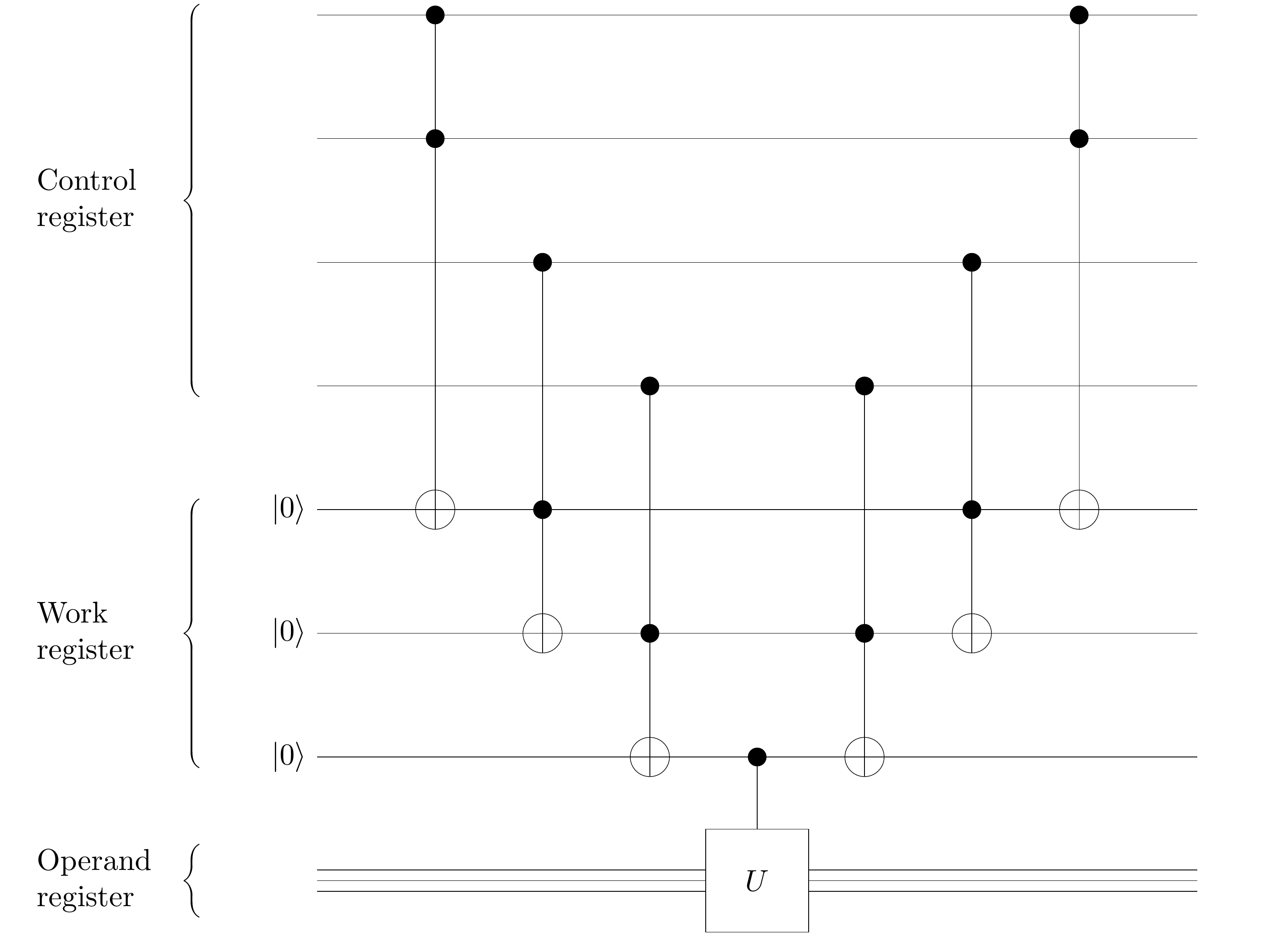}
    \caption{Control procedure summary for a 4-qubit control register and arbitrary operator $U$.
             This approach provides $\mathcal{O}\left(n\right)$ complexity for the implementation of an $n$-qubit control.
             This circuit is only applicable for a $\ket{1111}$ control state, but other controls can be easily implemented by applying Pauli $X$ gates to qubits requiring a $\ket{0}$ control.
             Note that the actual application of $U$ involves only a single control qubit, and hence the complexity of $U$ does not compound with the control complexity.
             }
    \label{fig:Control}
\end{center}
\end{figure}
While not strictly an issue for a naive implementation, the manner in which controls are applied is important for efficiency.
Consider the $B_j$ operator.
If row $j$ of $A$ contains $N^j_{nz}$ nonzero elements, then $B_j$ must apply $N^j_{nz}$ ancilla rotations, each with an $n$-qubit control.
The default in-place control scheme provided by, for example, Qiskit requires time exponential in $n$ to apply this control, thereby destroying the efficiency of the procedure.
Reference \cite{Barenco} provides a method for implementing such controlled operations with $\mathcal{O}\left(n^2\right)$ complexity without including any additional qubits, 
but this approach requires $n$ recursive square roots of the basic unitary to be computed.
Even for unitaries where these square roots can be computed easily, 
the complexity of their implementation in terms of basic gates scales poorly as the precision needed to accurately describe the desired operator increases.
This could potentially be addressed by providing a constant cutoff precision, 
but here we instead opt for a control scheme of complexity $\mathcal{O}\left(n\right)$ described in \cite{Nielsen}.
A diagram summarizing the procedure is shown in Fig. \ref{fig:Control}.
This method has the downside that it requires an additional $\left(n-1\right)$-qubit ``work'' register to store intermediate calculations related to the control procedure.
While this increases the absolute size of system required to apply a solution circuit, it does not affect the overall resource scaling of the procedure, which remains $\mathcal{O}\left(n+n_p\right)$.
Additionally, it provides an optimal complexity scaling for the control procedure, and applies the desired unitary directly.

Using this control scheme, the rotation component of $B_j$ will require $\mathcal{O}\left(N^j_{nz}\log\left(N\right)\right)$ basic gates to implement.
The initial Hadamard step to produce the uniform superposition can clearly be implemented with $\mathcal{O}\left(\log\left(N\right)\right)$ gates, and the Pauli $X$ gate requires constant complexity.
Therefore, the total gate complexity of each $B_j$ remains $\mathcal{O}\left(N^j_{nz}\log\left(N\right)\right)$.
Each application of $T_0$ and $W$ requires $N$ controlled $B_j$ operations, each requiring an $\left(n+1\right)$-qubit control.
However, note that the actual application of $B_j$ requires only a single control qubit, and hence the complexity of the control procedure is separate from that of $B_j$.
Then a single controlled $B_j$ has $\mathcal{O}\left(\log\left(N\right)\right)+\mathcal{O}\left(N^j_{nz}\log\left(N\right)\right)=\mathcal{O}\left(N^j_{nz}\log\left(N\right)\right)$ gate complexity.
Note that we have assumed $N^j_{nz}\ge 1$ for all $j$, which must always be satisfied for a well-conditioned system.
For constant precision, the solution circuit will require a constant number of applications of $W$ and $T_0$.
Then, noting that $\sum_{j=0}^{N-1}N^j_{nz}=N_{nz}$, this gives the final gate complexity of the solution procedure:
\begin{equation}
    \mathcal{O}\left(N_{nz}\log\left(N\right)\right)
\end{equation}

The last point of concern is the new procedure for eigenvalue extraction.
Where the canonical HHL unitary gives the eigenphase relationship $e^{2\pi i\phi_j}=e^{i\lambda_j}$, the walk unitary has $e^{2\pi i\phi_j}=i\hat{\lambda}_j\pm\sqrt{1-\hat{\lambda}_j^2}$, and hence the extraction procedure must be modified.
Recalling that $\hat{\lambda}_j\in\left[-1,1\right]$, this extraction is straightforward if we take the imaginary components:
\begin{equation}\label{eq:evalextract1}
    \sin\left(2\pi\phi_j\right) = \hat{\lambda}_j \implies \lambda_j = X\sin\left(2\pi\phi_j\right).
\end{equation}

\section{Treatment of Arbitrary Matrix Equations}\label{sec:Arbitrary}
Recall the following list of restrictions that have been imposed on our system:
\begin{enumerate}
    \item $\ket{b}$ must be normalized.
    \item $A$ must be Hermitian.
    \item $N$ must be a power of two.
    \item\label{enum:evals} The eigenvalues of $A$ must lie on the interval $\left[-X,X\right]$.
    \item No negative real numbers may appear on the diagonal of $A$.
\end{enumerate}
In this section, we develop a method by which an arbitrary matrix equation can be modified to satisfy these constraints.
To keep our notation consistent, we begin from the following initial problem: 
Given $A_0\in\mathbb{C}^{M\times M}$ and $\ket{b_0}\in\mathbb{C}^{M}$, find $\ket{x_0}$ such that
\begin{equation}\label{eq:sysprep_init}
    A_0\ket{x_0}=\ket{b_0}.
\end{equation}
This arbitrary input system provides the foundational elements for an appropriately restricted system.

The first restriction is that $\ket{b}$ is normalized.
This is easily satisfied by dividing (\ref{eq:sysprep_init}) by the magnitude of $\ket{b_0}$:
\begin{equation}
    \frac{1}{\left|\left|b_0\right|\right|_2} A_0\ket{x_0}=\frac{1}{\left|\left|b_0\right|\right|_2}\ket{b_0}.
\end{equation}
Here we assume that $\left|\left|b_0\right|\right|_2\ne 0$.
Since the zero vector is always a valid solution when $\left|\left|b_0\right|\right|_2 = 0$, this case is trivial and need not be considered for our purposes.

The second assumption is that the system matrix $A$ must be Hermitian.
This is satisfied in general by expanding the problem to the following $2M\times 2M$ system:
\begin{equation}
    \frac{1}{\left|\left|b_0\right|\right|_2}
    \begin{bmatrix}
        0 & A_0 \\
        A_0^{\dagger} & 0
    \end{bmatrix}
    \begin{bmatrix}
        0 \\ x_0
    \end{bmatrix}
    =
    \frac{1}{\left|\left|b_0\right|\right|_2}
    \begin{bmatrix}
        b_0 \\ 0
    \end{bmatrix}.
\end{equation}

Third, we consider that the size of our quantum system must be a power of 2, as the algorithm is constructed for quantum systems using two-state qubits.
Therefore, we define $n=\lceil\log\left(2M\right)\rceil$ and $N=2^n$, and once again expand our system, this time to
\begin{equation}\label{eq:sysprep_exp2}
    \frac{1}{\left|\left|b_0\right|\right|_2}
    \begin{bmatrix}
        0 & A_0 & 0 \\
        A_0^{\dagger} & 0 & 0 \\
        0 & 0 & I_{N-2M}
    \end{bmatrix}
    \begin{bmatrix}
        0 \\ x_0 \\ 0
    \end{bmatrix}
    =
    \frac{1}{\left|\left|b_0\right|\right|_2}
    \begin{bmatrix}
        b_0 \\ 0 \\ 0
    \end{bmatrix},
\end{equation}
where $I_{N-2M}$ is the identity matrix of dimension $N-2M$, and we have appended $N-2M$ zeros to each vector.
Now each vector can be represented by an $n$-qubit state.

The final problem is the restriction of the eigenvalues of $A$ to the range $\left[-X,X\right]$.
Let $\lambda_{\text{max}}$ be the magnitude of the dominant eigenvalue of $A$.
Then we have
\begin{equation}
    \lambda_{\text{max}} \le N\left|A_{jk}\right|_{\text{max}} \le X,
\end{equation}
by the discussion following (\ref{eq:PreppedState}).
Thus $\left[-\lambda_{\text{max}},\lambda_{\text{max}}\right]\subseteq\left[-X,X\right]$, and hence the spectrum of $A$ lies entirely on the required range.
That is, for this procedure, no rescaling of the system is required to satisfy the eigenvalue restrictions.

Every element on the diagonal of the system matrix is now either 0 or a positive real number, and hence no negative real elements appear on its diagonal.
This accounts for all imposed restrictions, leaving us with the following scheme for the preparation of an appropriate system:
\begin{align}
    &A = 
    \frac{1}{\left|\left|b_0\right|\right|_2}
    \begin{bmatrix}
        0 & A_0 & 0 \\
        A_0^{\dagger} & 0 & 0 \\
        0 & 0 & I_{N-2M}
    \end{bmatrix}
    , \nonumber \\
    &\ket{b} =
    \frac{1}{\left|\left|b_0\right|\right|_2}
    \begin{bmatrix}
        b_0 \\ 0 \\ 0
    \end{bmatrix}
    , \nonumber \\
    &\ket{x} =
    \begin{bmatrix}
        0 \\ x_0 \\ 0
    \end{bmatrix}.
\end{align}

Of course, it is possible that $A_0$ is already Hermitian, in which case some efficiency can be gained by not performing the full expansion as stated above.
Instead, only the size restriction must be satisfied.
To this end, we redefine $n=\lceil\log\left(M\right)\rceil$---keeping the definition $N=2^n$---and expand the system as follows:
\begin{align}
    &A=
    \frac{1}{\left|\left|b_0\right|\right|_2}
    \begin{bmatrix}
        A_0 & 0 \\
        0 & I_{N-M}
    \end{bmatrix},
    \nonumber \\
    &\ket{b}
    =
    \frac{1}{\left|\left|b_0\right|\right|_2}
    \begin{bmatrix}
        b_0 \\ 0
    \end{bmatrix},
    \nonumber \\
    &\ket{x}
    =
    \begin{bmatrix}
        x_0 \\ 0
    \end{bmatrix}.
\end{align}
Now it is possible for the system matrix to have negative real values on the diagonal.
This can be amended by using a shifted system:
\begin{align}
    \left(A+dI\right)\ket{x} = \ket{b},
\end{align}
where $d$ is an upper bound on the magnitude of the offending values on the diagonal of $A$.
For the purposes of the walk operator, $A+dI$ should be used as the system matrix, but note that this has the effect of shifting all eigenvalues of $A$ by $d$.
Hence, in order to directly apply the inverse of $A$ to the initial state, the eigenvalue extraction of (\ref{eq:evalextract1}) must be modified:
\begin{equation}\label{eq:evalextract2}
    \lambda_j = X\sin\left(2\pi\phi_j\right) - d.
\end{equation}
So long as $X$ is calculated from the shifted system matrix, the eigenvalues are still appropriately bounded, and $A$ still requires no additional rescaling to satisfy the eigenvalue restrictions.

\section{Summary}
\begin{figure*}[ht!]
\begin{center}
    \includegraphics[width=\textwidth]{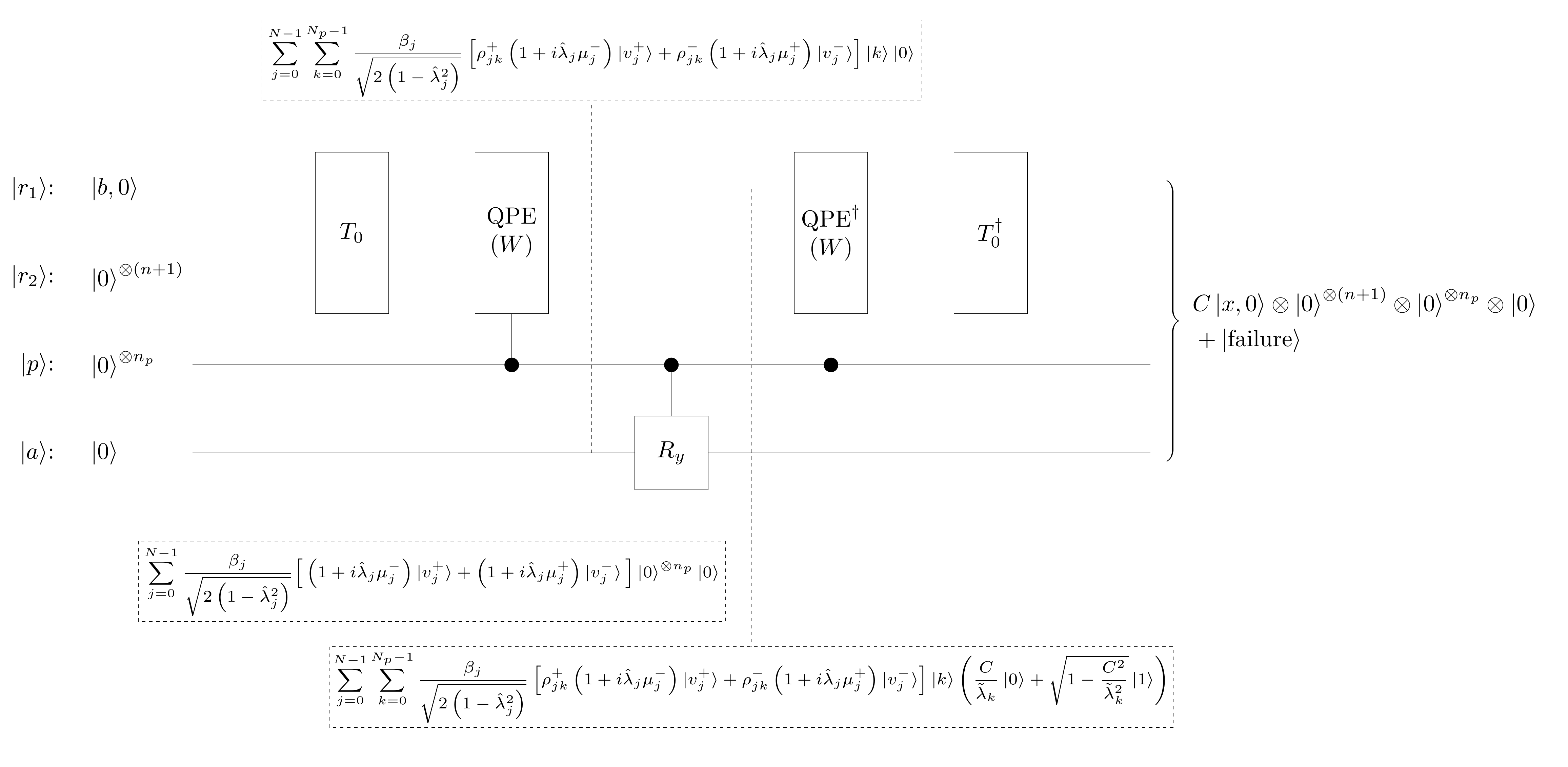}
    \caption{Structural outline of the full procedure.
             Dashed boxes indicate the state of the system after every major intermediate stage.
             The $\ket{r_1}$ and $\ket{r_2}$ registers are the ancilla-augmented vector registers on which the walk operator and its associated constituents act.
             The phase register $\ket{p}$ provides space for the estimation of the eigenphases of $W$.
             The ancilla $\ket{a}$ is used for the application of the HHL rotation described in section \ref{sec:HHL}.
             We commonly refer to $\ket{a}$ as the ``HHL ancilla'' to distinguish it from the ancillae of $\ket{r_1}$ and $\ket{r_2}$.
             The phase amplitudes $\rho_{jk}^+$ and $\rho_{jk}^-$ are used to distinguish the amplitudes of phase estimates corresponding to each of the two eigenvectors $\ket{v_j^{\pm}}$.
             The eigenvalue estimates themselves require no such distinction, as both $\mu_j^+$ and $\mu_j^-$ have the same imaginary part.
             }
    \label{fig:overview}
\end{center}
\end{figure*}
Here we summarize the preceding analysis in the form of a self-contained procedure.
We assume that the supplied matrix equation has been properly prepared according to section \ref{sec:Arbitrary}.
Fig. \ref{fig:overview} provides a large-scale pictorial description of the full procedure.
Notice that this procedure consists of four main stages: the initial application of $T_0$, quantum phase estimation, HHL rotation, and uncomputation.

\subsection{Initial Application of $T_0$}
\begin{figure*}
\begin{center}
    \includegraphics[width=\textwidth]{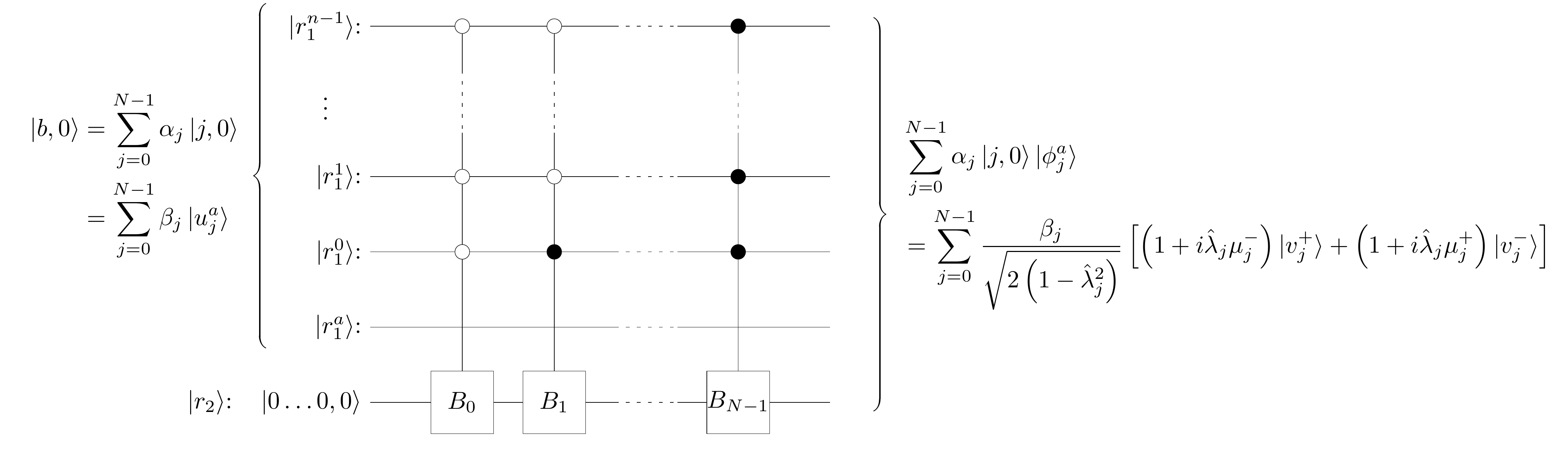}
    \caption{Diagram of the initial application of $T_0$.
             Here state preparation operators are applied to $\ket{r_2}$, conditioned on the state of $\ket{r_1}$.
             Note that, due to the presumption that $\ket{r_1}$ is initially prepared with an ancilla state of $\ket{0}$, the $\ket{r_1}$ ancilla is ignored in this calculation.
             Since the resulting system is highly entangled, individual qubit states cannot be separated.
             As a result, we have only provided a final state for the total system.
             }
    \label{fig:T0}
\end{center}
\end{figure*}
\begin{figure*}
\begin{center}
    \includegraphics[width=\textwidth]{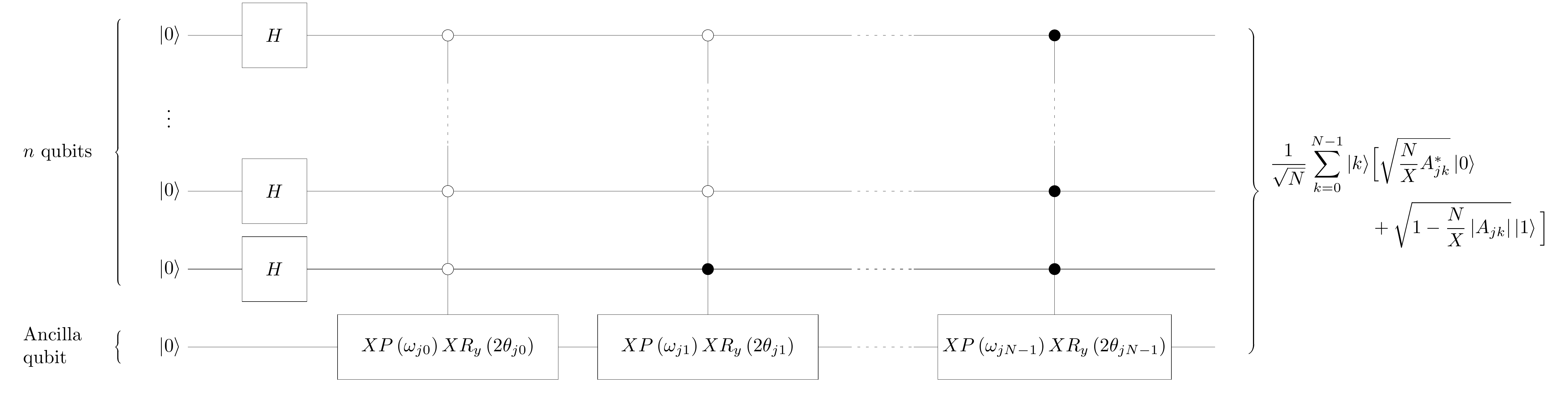}
    \caption{Diagram of the state preparation operator $B_j$ applied to a register with all qubits initialized to the $\ket{0}$ state.
             Here $\theta_{jk}=\arccos\left(\sqrt{\left|A_{jk}\right|N/X}\right)$ and $\omega_{jk}=\pm\arg\left(A_{jk}\right)/2$, where the positive solution for $\omega_{jk}$ is chosen iff $\arg\left(A_{jk}\right)=\pi$ and $j>k$.
             Thus, the $R_y$ gates rotate according to the magnitudes of the elements of $A$, with the factor of $XPX$ subsequently enforcing the correct phase on the $\ket{0}$ state of the ancilla.
             }
    \label{fig:Bj}
\end{center}
\end{figure*}
\begin{algorithm}
\caption{Implementation of $T_0$}\label{alg:T0}
\begin{algorithmic}[1]
    \For{$j=0 \dots N-1$}
        \State Compute the operator $B_j$
        \State Apply $B_j$ to $\ket{r_2}$ with the control condition that $\ket{r_1}$
        \Statex \hspace{\algorithmicindent}is in the state $\ket{j,0}$
    \EndFor
\end{algorithmic}
\end{algorithm}
\begin{algorithm}
\caption{Implementation of $B_j$}\label{alg:Bj}
\begin{algorithmic}[1]
    \For{$k=0 \dots n-1$}
       \State Apply $H$ to qubit $k$
    \EndFor
    \For{$k=0 \dots N-1$}
        \State $\theta_{jk}\leftarrow\arccos\left(\sqrt{\left|A_{jk}\right|N/X}\right)$
        \State $\omega_{jk}\leftarrow-\arg\left(A_{jk}\right)/2$
        \If{$\arg\left(A_{jk}\right)=\pi$ and $j<k$}
            \State $\omega_{jk}\leftarrow-\omega_{jk}$
        \EndIf
        \State Apply $XP\left(\omega_{jk}\right)XR_y\left(2\theta_{jk}\right)$ to the ancilla qubit
        \Statex \hspace{\algorithmicindent}with the control condition that the rest of the register
        \Statex \hspace{\algorithmicindent}is in the state $\ket{k}$
    \EndFor
\end{algorithmic}
\end{algorithm}
The first stage of the procedure is the application of $T_0$, which maps $\ket{r_2}$ to a superposition of the
$\ket{\phi_j^a}$ and $\ket{\zeta_j^a}$ states defined in section \ref{sec:WalkOp}.
A circuit diagram for our suggested implementation of $T_0$ is shown in Fig. \ref{fig:T0}, and a pseudocode description of the operator is given in algorithm \ref{alg:T0}.
In the program, the ``T0'' function of the ``QWOps'' module implements this operator.
The definition of $T_0$ depends on the state preparation operators $B_j$, and 
as such we have also provided a circuit diagram for an arbitrary $B_j$ in Fig. \ref{fig:Bj}.
The corresponding pseudocode description is given in algorithm \ref{alg:Bj}.
In the program, $B_j$ is implemented by the ``Bj'' function of the ``QWOps'' module.
Here we have chosen to forgo any application of the $B_j^{\prime}$ operator, as we can safely presume that the ancilla state of $\ket{r_1}$ will in practice be initialized to $\ket{0}$ .

In this application of $T_0$, the system undergoes the transformation
\begin{align}
    &\ket{b,0}\ket{0\dots0,0} = \sum_{j=0}^{N-1} \beta_j\ket{u_j,0}\ket{0\dots0,0}
    \nonumber \\
    \overset{T_0}{\longmapsto}&
    \sum_{j=0}^{N-1} \frac{\beta_j}{\sqrt{2\left(1-\hat{\lambda}_j^2\right)}} \Big[\left(1+i\hat{\lambda}_j\mu_j^-\right)\ket{v_j^+} 
    \nonumber \\
    &\hphantom{\sum_{j=0}^{N-1} \frac{\beta_j}{\sqrt{2\left(1-\hat{\lambda}_j^2\right)}}\Big[} + \left(1+i\hat{\lambda}_j\mu_j^+\right)\ket{v_j^-}\Big].
\end{align}
Here we have suppressed the phase and HHL ancilla registers, as they are not subject to the effects of $T_0$.

\subsection{Quantum Phase Estimation}
\begin{figure}
\begin{center}
    \includegraphics[width=\columnwidth]{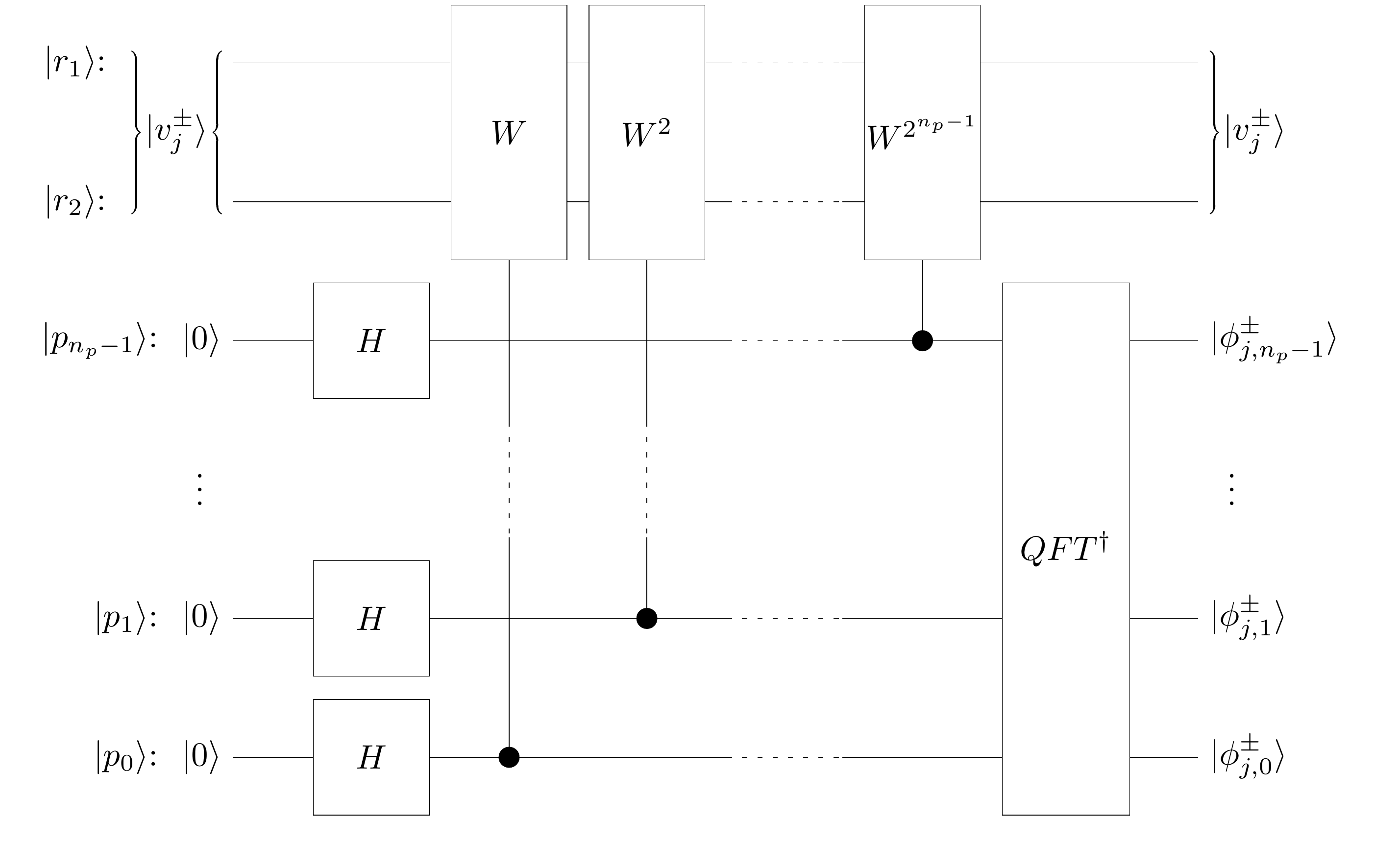}
    \caption{Diagram of phase estimation using the walk operator.
             Here we use the same simplifications as in Fig. \ref{fig:QPE-basic}.
             }
    \label{fig:QPE}
\end{center}
\end{figure}

\begin{figure*}
\begin{center}
    \includegraphics[width=\textwidth]{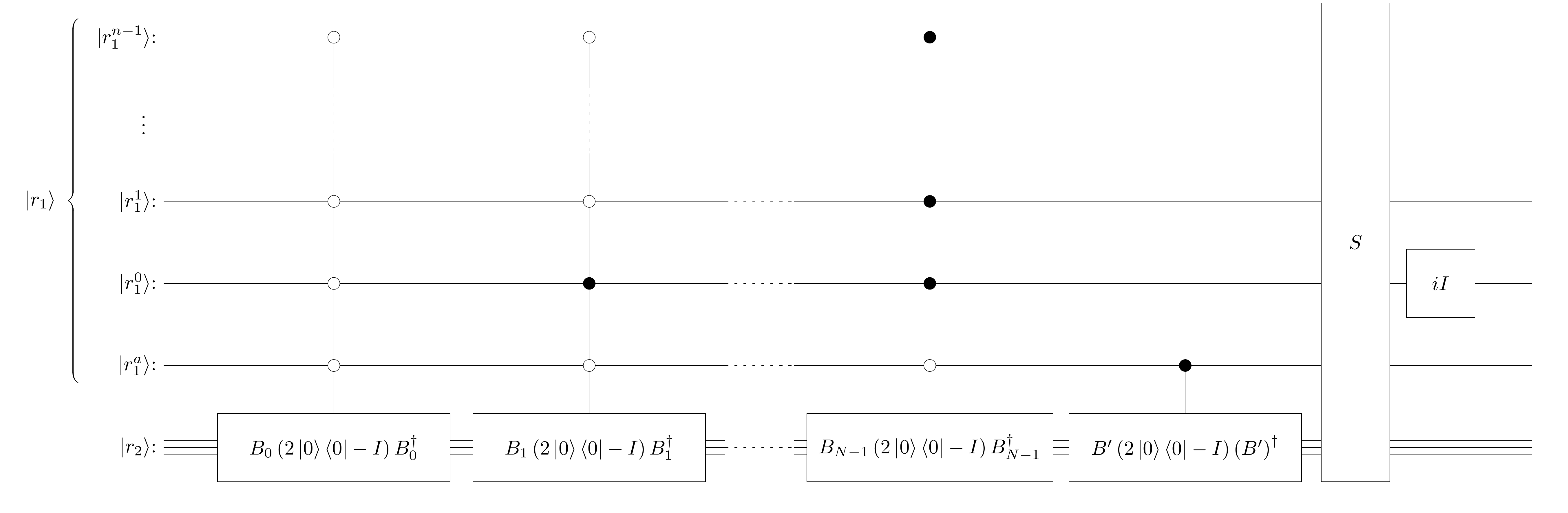}
    \caption{Diagram of the walk operator $W$.
             Here $S$ represents a one-to-one exchange of qubits states between $\ket{r_1}$ and $\ket{r_2}$.
             The operator $iI$ can be applied to any qubit in $\ket{r_1}$ or $\ket{r_2}$ to produce the desired effect.
             }
    \label{fig:W}
\end{center}
\end{figure*}

With $\ket{r_1}$ and $\ket{r_2}$ prepared in the desired superposition of the walk operator's eigenvectors, an application of QPE using $W$ as the unitary writes the eigenphases of $W$ onto the phase register $\ket{p}$.
An expansion of this step, in the typical QPE format, is given in Fig. \ref{fig:QPE}.
A circuit diagram for the walk operator itself is shown in Fig. \ref{fig:W}, and the corresponding pseudocode is given in algorithm \ref{alg:W}.
The walk operator is implemented in the program by the ``W'' function of the ``QWOps'' module.
Note that we have made use of our suggested $j$-independent form of the $\ket{\zeta_j^a}$ states, wherein $\ket{\zeta_j^a}=\ket{0}^{\otimes n}\ket{1}$.
This eliminates the $j$ dependence of the corresponding preparation operator, which we now refer to as simply $B^{\prime}$.
Since the swap operations can leave important data on the $\ket{1}$ state of $\ket{r_1}$'s ancilla, $B^{\prime}$ cannot be ignored as in the initial application of $T_0$.
In the model of our current study, $B^{\prime}$ can be implemented by simply applying an $X$ gate to the ancilla of $\ket{r_2}$.
The program implements $B^{\prime}$ in the ``Bp'' function of the ``QWOps'' module.

After phase estimation, the system is in the state
\begin{align}
    &\sum_{j=0}^{N-1}\sum_{k=0}^{N_p-1} \frac{\beta_j}{\sqrt{2\left(1-\hat{\lambda}_j^2\right)}}
    \nonumber \\
    &\hphantom{\sum_{j=0}^{N-1}} \Big[\rho_{jk}^+\left(1+i\hat{\lambda}_j\mu_j^-\right)\ket{v_j^+} + \rho_{jk}^-\left(1+i\hat{\lambda}_j\mu_j^+\right)\ket{v_j^-}\Big]\ket{k}.
\end{align}
Here we have reintroduced the phase register, although $\ket{a}$ remains suppressed.
As in section \ref{sec:HHL}, $\rho_{jk}^{\pm}$ extracts accurate eigenvalue approximations by spiking sharply when $\tilde{\lambda}_k$ approaches an actual eigenvalue $\lambda_j$.

\subsection{HHL Ancilla Rotation}
\begin{figure*}
\begin{center}
    \includegraphics[width=\textwidth]{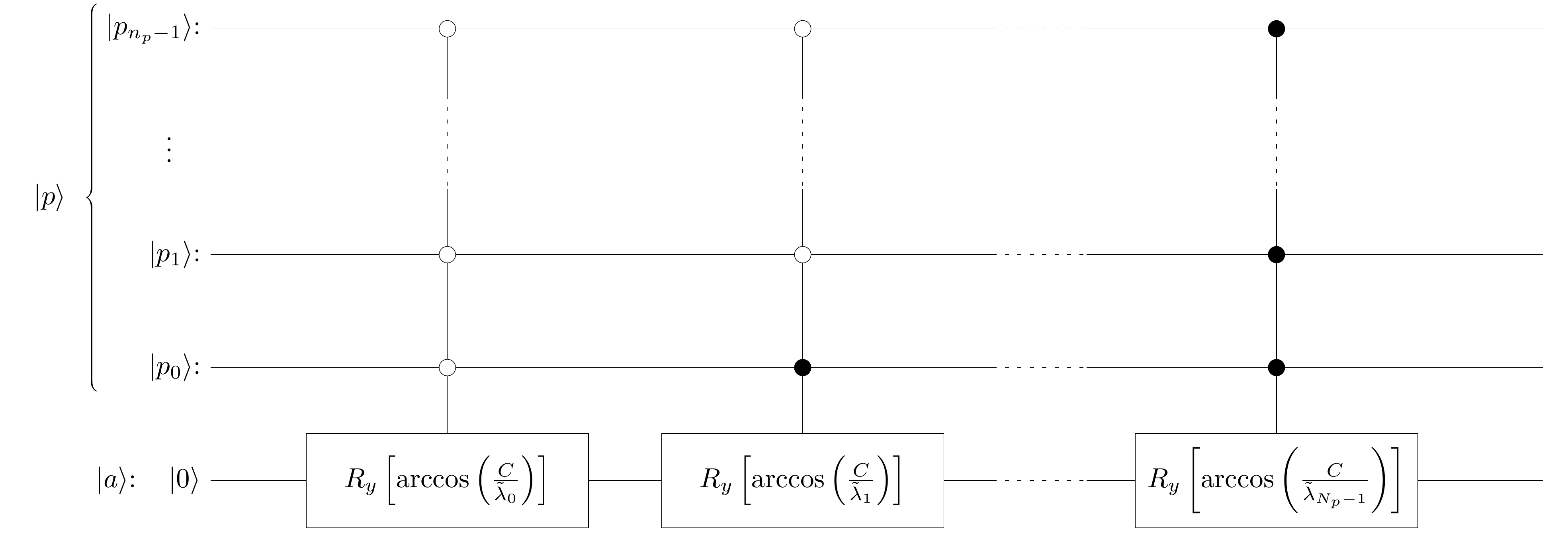}
    \caption{Diagram of the HHL rotation procedure.
             Here $\tilde{\lambda}_k = X\sin\left(2\pi k/N_p\right) - d$, and $C$ is a lower bound on the magnitude of the $\tilde{\lambda}_k$'s that are not zero.
             Only those rotations for which $\tilde{\lambda}_k\ne 0$ should be applied.
             }
    \label{fig:Rot}
\end{center}
\end{figure*}
\begin{algorithm}
\caption{Implementation of the HHL rotation}\label{alg:Rot}
\begin{algorithmic}[1]
    \For{$k=0 \dots N_p-1$}
        \State $\tilde{\lambda}_k \leftarrow X\sin\left(2\pi k/N_p\right) - d$
        \If{$\tilde{\lambda}_k\ne0$}
            \State Apply the rotation $R_y\left[2\arccos\left(C/\tilde{\lambda}_k\right)\right]$ to $\ket{a}$, 
            \Statex \hspace{\algorithmicindent}\hspace{\algorithmicindent}with the control condition that $\ket{p}$ is in the state
            \Statex \hspace{\algorithmicindent}\hspace{\algorithmicindent}$\ket{k}$
        \EndIf
    \EndFor
\end{algorithmic}
\end{algorithm}
With the phase register holding the eigenphases of $W$, extraction of the corresponding eigenvalues can proceed according to the analysis of sections \ref{sec:Practical} and \ref{sec:Arbitrary}.
Specifically, the eigenvalues are calculated according to
\begin{equation}
    \lambda_j = X\sin\left(2\pi\phi_j\right) - d,
\end{equation}
where $d=0$ can be used if the system was expanded to satisfy the Hermitian system matrix property.
Of course, the exact values of $\phi_j$ are not known in general, and in practice approximate eigenvalues must be calculated from the state of the phase register as follows:
\begin{equation}
    \tilde{\lambda}_k = X\sin\left(2\pi\frac{k}{N_p}\right)-d.
\end{equation}
Once these eigenvalue approximations have been calculated, a rotation operator can be used to apply an eigenvalue factor corresponding to the inverse of the system matrix for every possible state of the phase register, as detailed in section \ref{sec:HHL}.
A circuit diagram for this procedure is given in Fig. \ref{fig:Rot}, and the relevant pseudocode is provided in algorithm \ref{alg:Rot}.
This rotation is implemented in the program by the ``HHLRotation'' function of the ``QAlgs'' module.

\begin{algorithm}
\caption{Implementation of $W$}\label{alg:W}
\begin{algorithmic}[1]
    \For{$j=0 \dots N-1$}
        \State Compute the state preparation operator $B_j$
        \State Apply the reflector $B_j\left(2\ket{0}\bra{0}-I\right)B_j^{\dagger}$ to $\ket{r_2}$ with 
        \Statex \hspace{\algorithmicindent}the control condition that $\ket{r_1}$ is in the state $\ket{j,0}$
    \EndFor
    \State Compute the state preparation operator $B^{\prime}$
    \State Apply the reflector $B^{\prime}\left(2\ket{0}\bra{0}-I\right)\left(B^{\prime}\right)^{\dagger}$ to $\ket{r_2}$ with 
    \Statex the control condition that the $\ket{r_1}$ ancilla is in the state
    \Statex $\ket{1}$
    \For{$j=0 \dots N-1$}
        \State Apply a swap operation to the $j$-th qubits of $\ket{r_1}$ and
        \Statex \hspace{\algorithmicindent}$\ket{r_2}$
    \EndFor
    \State Apply a swap operation to the ancillae of $\ket{r_1}$ and $\ket{r_2}$
    \State Apply $iI$ to any qubit of $\ket{r_1}$ or $\ket{r_2}$
\end{algorithmic}
\end{algorithm}

Upon completion of this rotation, the system has the state
\begin{align}
    \sum_{j=0}^{N-1}&\sum_{k=0}^{N_p-1} \frac{\beta_j}{\sqrt{2\left(1-\hat{\lambda}_j^2\right)}}
    \nonumber \\
    &\Big[\rho_{jk}^+\left(1+i\hat{\lambda}_j\mu_j^-\right)\ket{v_j^+} + \rho_{jk}^-\left(1+i\hat{\lambda}_j\mu_j^+\right)\ket{v_j^-}\Big]\ket{k}
    \nonumber \\
    \otimes&\left(\frac{C}{\tilde{\lambda}}_k\ket{0} + \sqrt{1-\frac{C^2}{\tilde{\lambda}_k^2}}\ket{1}\right).
\end{align}

\subsection{Uncomputation}
After rotation, $\ket{p}$, $\ket{r_2}$, and the ancilla of $\ket{r_1}$ are uncomputed by applying inverse QPE and $T_0$ operations.
Ideally, these registers are returned to the $\ket{0}$ state, and any other state can be interpreted as a failure of the procedure.
This leaves $\ket{r_1}$ entangled with $\ket{a}$. 
When $\ket{a}$ takes the state $\ket{0}$, the state of $\ket{r_1}$ is proportional (up to the error introduced through QPE) to the solution state $\ket{x}$.
Meanwhile, if $\ket{a}$ is in the state $\ket{1}$, the system is in a failure state.
That is, the overall final state of the system is
\begin{equation}
    C\ket{x,0}\ket{0}^{\otimes \left(n+1\right)}\ket{0}^{\otimes n_p}\ket{0} + \ket{\text{failure}}.
\end{equation}
This final state is analogous to (\ref{eq:HHL_res}).

From this description, we see that the full procedure is well defined for any input matrix, with all constituent operations having explicit representations.
This is in contrast to the canonical HHL algorithm, where the implementation of the QPE unitary presents an ambiguously defined bottleneck for the procedure.

\section{Results}\label{sec:Results}
In developing suitable tests for the practical performance of this procedure, two important limitations must be considered:
First, the quantum systems currently available to the public are quite small, with the largest system we have access to offering only 7 qubits\footnote{This is the \emph{ibmq\_jakarta} v1.0.23 system, 
which is one of the IBM Quantum Falcon Processors.}\cite{IBMQ}.
Second, errors on these systems can propagate rapidly, with calculations involving more than ${\sim}800$ primitive gates being, in our experience, unlikely to yield the expected states.
Therefore, when considering a problem suitable for fully quantum analysis, we are at present limited to very short, small calculations.
Section \ref{ssec:Quantum} constructs and analyzes such a problem, but it is possible to construct a more satisfying test using classical simulation.
By using a classical computer to simulate a quantum system, we gain access to more qubits and eliminate gate errors.
Such simulation is extremely inefficient, but we have found that as many as 14 qubits can be simulated for this procedure using available classical systems.
Section \ref{ssec:Simulated} considers a larger, less restricted test problem using classical simulation.

Aside from verifying the accuracy of our method, we must also show that it is efficient compared to classical algorithms.
For this task, we use the example of sparse approximate inverse (SPAI) preconditioning.
Section \ref{ssec:Preconditioner} is dedicated analyzing the efficiency of preconditioner application.

\subsection{A Fully Quantum Problem}\label{ssec:Quantum}
Being limited to 7 qubits, we must consider only the smallest of possible problems.
Thus we choose a problem of two unknowns, requiring a single qubit for each of the main portions of $\ket{r_1}$ and $\ket{r_2}$.
Both of these registers require ancillary qubits, and a third ancilla is also required for the eigenvalue rotation.
Then we currently stand at 5 qubits accounted for, with 2 left for the phase register.
Since we are most concerned with using a minimal number of qubits, we use the exponential time, in-place control scheme provided by default in Qiskit.
Thus, no additional qubits are needed for control considerations.
These four possible phase states admit no error in the phase estimation, and hence we must choose a problem such that the eigenphases can be exactly represented by two bits.

By (\ref{eq:evalextract2}), the eigenvalues of the system matrix must satisfy
\begin{equation}
    \lambda_j = X\sin\left(2\pi\phi_j\right)-d
    = \begin{cases} -d,&\phi_j=0,\frac12 \\ X-d,&\phi_j=\frac14 \\ -X-d,&\phi_j=\frac34. \end{cases}
\end{equation}
The system
\begin{equation}
    A = \begin{bmatrix} -2 & 1 \\ 1 & -2 \end{bmatrix},
    \quad d=3, \quad X=2,
\end{equation}
satisfies the constraints of our procedure, and has eigenvalues
\begin{equation}
    \lambda_0 = -3 = -d, \quad \lambda_1 = -1 = X-d.
\end{equation}
Thus, this system provides a suitable candidate for our analysis.
Note that the system is Hermitian, and no expansion is required.
For a right-hand side vector, we use an equal superposition of both eigenvectors of $A$:
\begin{equation}
    \ket{b} = \frac12 \left(\begin{bmatrix}-1\\1\end{bmatrix}+\begin{bmatrix}1\\1\end{bmatrix}\right) = \begin{bmatrix}0\\1\end{bmatrix}.
\end{equation}

For this problem of known form, some simplification is possible in the various operators of the solution procedure.
This allows the size of the final circuit to be significantly reduced.
The system matrix defining the walk operator is
\begin{equation}
    A+dI = \begin{bmatrix} 1 & 1 \\ 1 & 1 \end{bmatrix}.
\end{equation}
Therefore, all $B_j$ operators apply the same ancilla rotation:
\begin{equation}
    R_y\left(2\arccos\sqrt{1\cdot\frac{2}{2}}\right)
    = R_y\left(0\right)
    = I.
\end{equation}
Since all elements are real, no phase rotation is necessary, and all $B_j$ operators can be implemented by simply applying a Hadamard gate to $\ket{r_2}$.
This elimination of the complicated phase and $R_y$ gates, both of which also require additional control considerations, greatly reduces the complexity of each $B_j$ operator.

This simplification of the $B_j$ operators carries through the $T_0$ and $W$ operators.
Without this reduction, the full program produces a circuit consisting of 15,728 basic gates.
This figure results from transpilation using the available $CX$, $I$, $R_z$, $\sqrt{X}$, and $X$ basis gates.
After simplification, the final circuit requires only 2,696 gates.
This is a notable reduction, but the circuit is still much too large considering the errors incurred by each gate application.
We remedy this by dividing the circuit into many smaller sub-circuits, and running each of these components as its own calculation.
By initializing each component based on a classically simulated result from the previous component (rather than continuing from an imprecise result as a direct calculation would), 
the compounding effect of gate errors can be eliminated.
We can then verify the validity of our solution by confirming that the results of each component agree with the simulated results for the same component.

\begin{figure*}
\begin{center}
    \begin{subfigure}{0.49\textwidth}
        \centering
        \includegraphics[width=\textwidth]{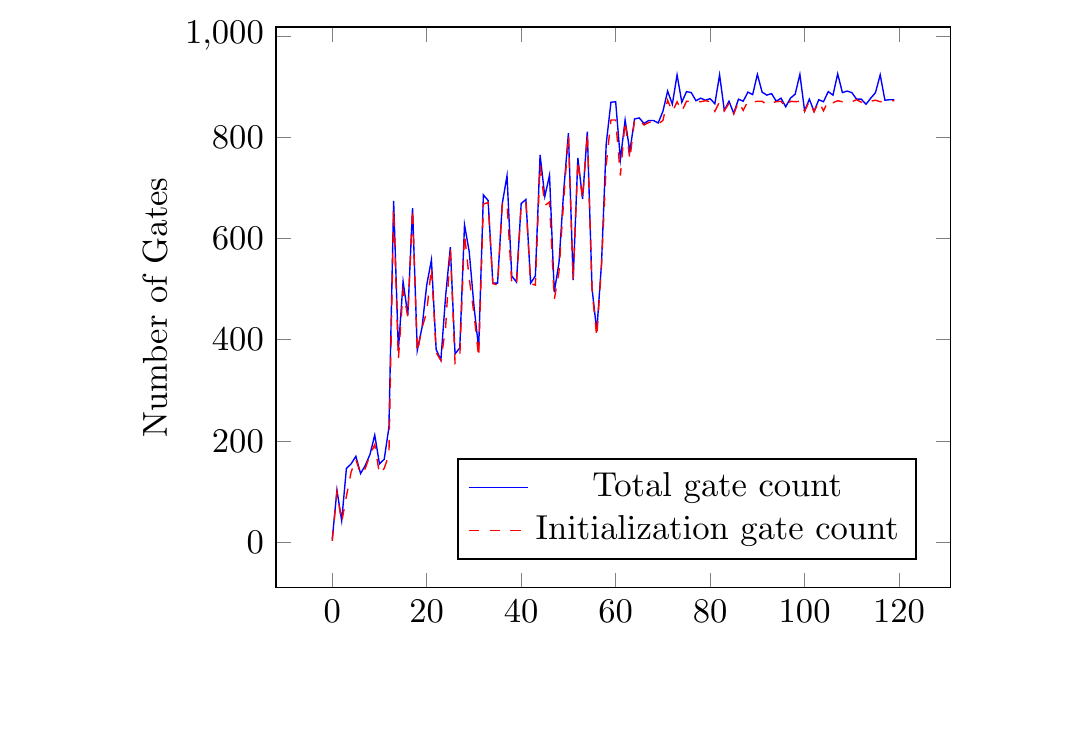}
        \caption{}
        \label{subfig:ComponentPlot}
    \end{subfigure}
    \begin{subfigure}{0.49\textwidth}
        \centering
        \includegraphics[width=\textwidth]{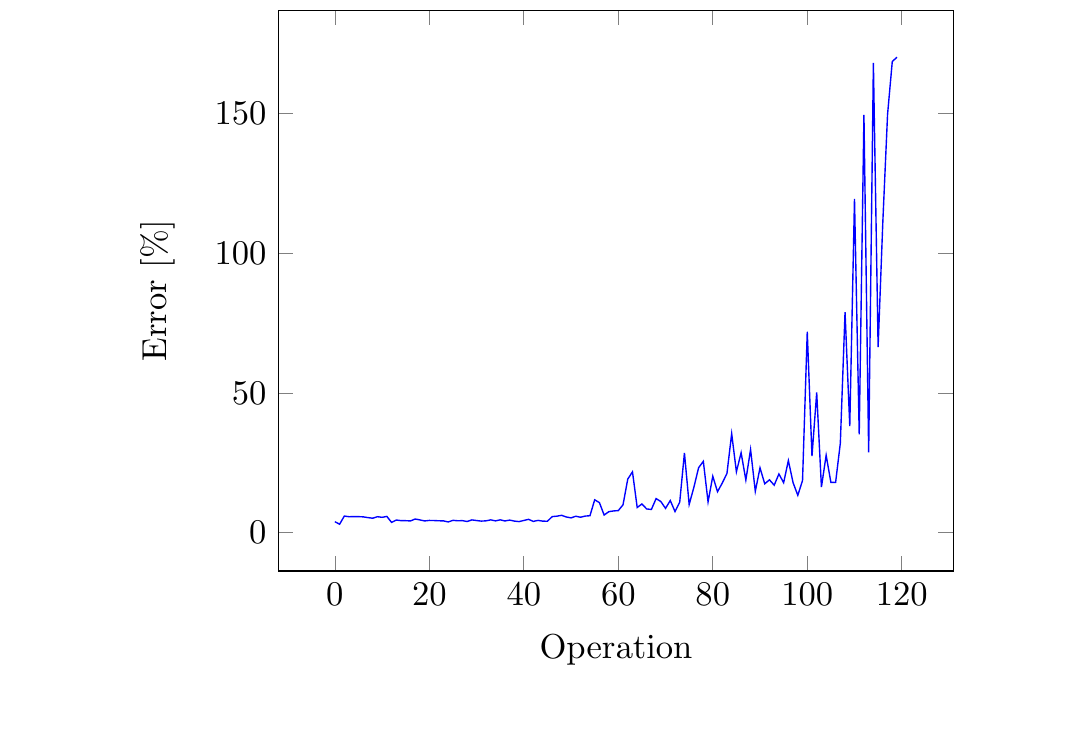}
        \caption{}
        \label{subfig:ErrorPropagation}
    \end{subfigure}
    \caption{Gate counts (\subref{subfig:ComponentPlot}) and errors (\subref{subfig:ErrorPropagation}) for each circuit component in the solution procedure.
             Note that, due to the complex intermediate states, most gates are dedicated to component initializations.}
    \label{fig:ComponentErrorPlot}
\end{center}
\end{figure*}
In total, the circuit was divided into 120 components as listed in appendix \ref{sec:Components}.
Initializations were performed using Qiskit's built-in initialization function using the state vectors produced by classical simulation of the previous components.
The results of the calculation are shown in Fig. \ref{fig:ComponentErrorPlot}.
We see that good agreement is maintained between the simulated and quantum procedures until the more complex later stages, when gate counts approach and exceed 800.
At this point, calculations quickly become unreliable.
Note that the sum of the number of gates required for all sub-circuits is significantly larger than the number of gates reported for the direct circuit.
This is due to the reinitialization of the system, which adds an unpredictable overhead cost to each operation.
The average relative error of the final simulated solution compared to a direct classical solution is $1.12\times10^{-10}$, which is well within the tolerances prescribed by the program.

\subsection{Simulated Solution}\label{ssec:Simulated}
With space available for a slightly larger system, we consider a practical, though still trivial, problem.
We consider a calculation of the charge per unit length on a transmission line consisting of two long, conducting strips of known potentials radiating in free space.
This problem can be modeled as a method of moments discretization of a cross-section of the line, yielding a matrix equation for the element charges $\ket{Q}$ as a function of their potentials $\ket{V}$:
\begin{equation}
    B\ket{Q}=\ket{V}.
\end{equation}
Elements of the matrix $B$ are given by \cite{Sadiku}:
\begin{equation}
    B_{jk} = \begin{cases} -\frac{l_j}{2\pi\varepsilon_0} \left[\ln\left(l_j\right)-1.5\right], & j=k \\ -\frac{l_j}{2\pi\varepsilon_0}\ln\left(d_{jk}\right), & j \ne k. \end{cases}
\end{equation}
where $l_j$ is the length of the $j$-th one-dimensional boundary element and $d_{jk}$ is the separation between the centroids of elements $j$ and $k$.
By using elements of uniform size $l$, we can ensure that $B$ is Hermitian.
We give each strip unit width, and orient both strips parallel to each other with unit separation.

\begin{figure}
\begin{center}
    \includegraphics[width=\columnwidth]{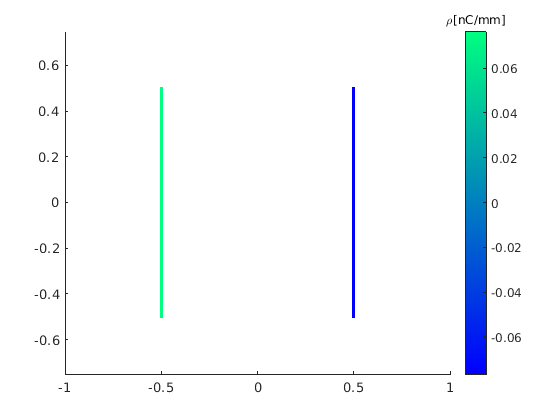}
    \caption{Solved charge distribution.
             The left conductor was held at a $1.0V$ potential, while the right conductor was held at $-1.0V$.
             Distances are shown in mm.
             }
    \label{fig:stripplot}
\end{center}
\end{figure}
\begin{table}[htb!]
\begin{center}
\caption{Comparison of Simulated Quantum and Direct Classical Solution Vectors}
\label{tab:solvecs}
\begin{tabular}{ |c|c|c|c| }
    \hline
    Element Index & Quantum $\left[nC\right]$ & Classical $\left[nC\right]$ & Relative Error\\
    \hline
    $0$ & $0.0383$ & $0.0371$ & $0.0315$ \\
    \hline
    $1$ & $0.0383$ & $0.0371$ & $0.0315$ \\
    \hline
    $2$ & $-0.0383$ & $-0.0371$ & $0.0315$ \\
    \hline
    $3$ & $-0.0383$ & $-0.0371$ & $0.0315$ \\
    \hline
\end{tabular}
\end{center}
\end{table}
The calculated charge density for each element is shown in Fig. \ref{fig:stripplot}, and the solution vector and its error are given in table \ref{tab:solvecs}.
For this calculation, four unknowns and seven phase qubits were used.
We see that the quantum results are within ${\sim}3\%$ of the correct solution, showing that our procedure produces a reliable approximation.
For more accurate solutions, a more complex system involving additional phase qubits would need to be simulated.

\subsection{Preconditioner Application}\label{ssec:Preconditioner}
\begin{figure}
\begin{center}
    \includegraphics[width=\columnwidth]{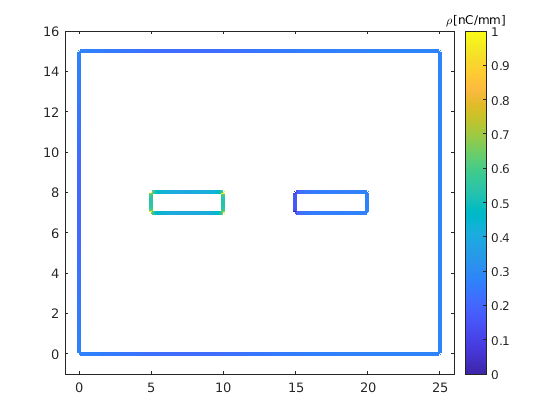}
    \caption{Classically solved charge distribution.
             Distances are shown in mm.}
    \label{fig:linedist}
\end{center}
\end{figure}
Here we consider the calculation of charge density per unit length on a two-conductor rectangular transmission line using the same method of moments approach as in the previous section.
For reference, a classical solution is shown in Fig. \ref{fig:linedist}.
This geometry is too complex for our quantum procedure to solve outright in satisfactory detail, 
but the larger density of charge elements provides a more enlightening example for preconditioning than the previous two-strip example.
To generate the preconditioner, we follow the same procedure as when generating the total system matrix, 
but only include elements corresponding to positions which lie within a certain distance $\delta$ of each other (we use four times the basic element size).
Explicitly, for preconditioner $P$,
\begin{equation}
    P_{jk} = \begin{cases} 
                 -\frac{l_j}{2\pi\varepsilon_0} \left[\ln\left(l_j\right)-1.5\right], & j=k \text{ and } d_{jk} \le \delta\\ 
                 -\frac{l_j}{2\pi\varepsilon_0}\ln\left(d_{jk}\right), & j \ne k \text{ and } d_{jk} \le \delta\\
                 0, & \text{otherwise}.
             \end{cases}
\end{equation}

\begin{figure*}
\begin{center}
    \begin{subfigure}{0.49\textwidth}
        \centering
        \includegraphics[width=\textwidth]{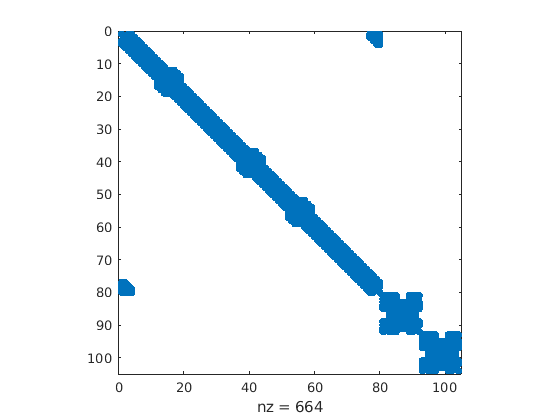}
        \caption{}
        \label{subfig:DensityP}
    \end{subfigure}
    \begin{subfigure}{0.49\textwidth}
        \centering
        \includegraphics[width=\textwidth]{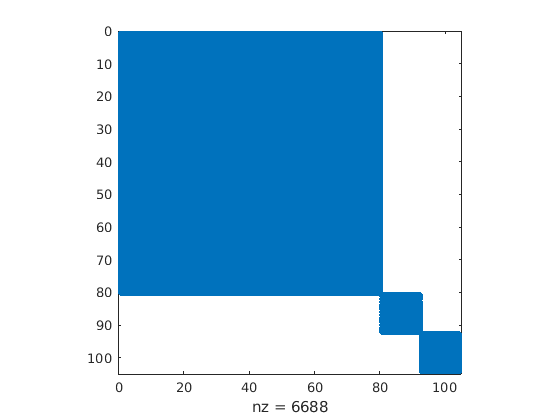}
        \caption{}
        \label{subfig:DensityPinv}
    \end{subfigure}
    \caption{Nonzeros in preconditioner $P$ (\subref{subfig:DensityP}) and $P^{-1}$ (\subref{subfig:DensityPinv}).}
    \label{fig:Density}
\end{center}
\end{figure*}
By applying $P^{-1}$ to the right-hand side vector of the matrix equation prior to performing an iterative solution procedure, the total number of iterations needed to solve the system can be drastically reduced \cite{Pan}.
However, even though $P$ is generally sparse, $P^{-1}$ can still be quite dense, as illustrated in Fig. \ref{fig:Density}.
This makes efficient classical application of the inverse preconditioner difficult, and often impossible.
However, we have already shown that our quantum procedure can apply $P$ in $\mathcal{O}\left(N_{nz}\log\left(N\right)\right)$ time, regardless of the sparsity of $P^{-1}$.

\begin{figure}
\begin{center}
    \includegraphics[width=\columnwidth]{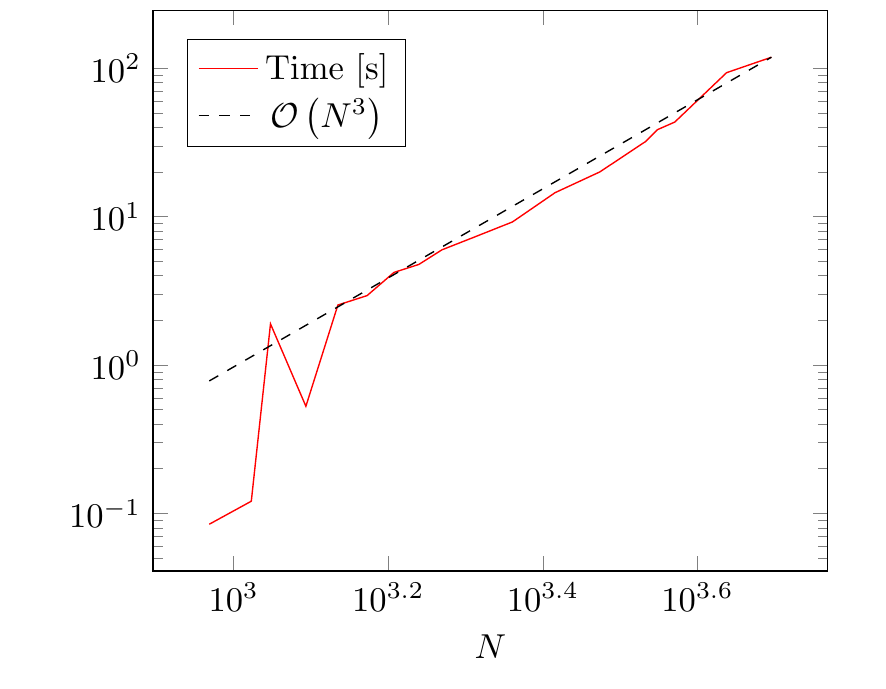}
    \caption{Time to apply inverse preconditioner classically using GMRES.}
    \label{fig:TimeScaleClassical}
\end{center}
\end{figure}
\begin{figure*}
\begin{center}
    \begin{subfigure}{0.49\textwidth}
        \centering
        \includegraphics[width=\textwidth]{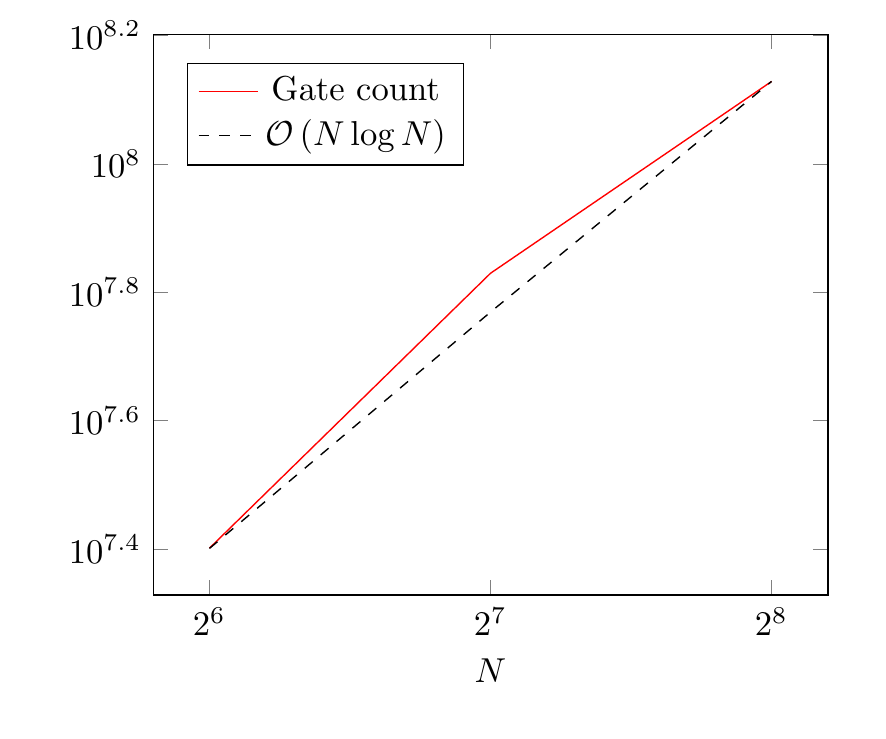}
        \caption{}
        \label{subfig:GateScale}
    \end{subfigure}
    \begin{subfigure}{0.49\textwidth}
        \centering
        \includegraphics[width=\textwidth]{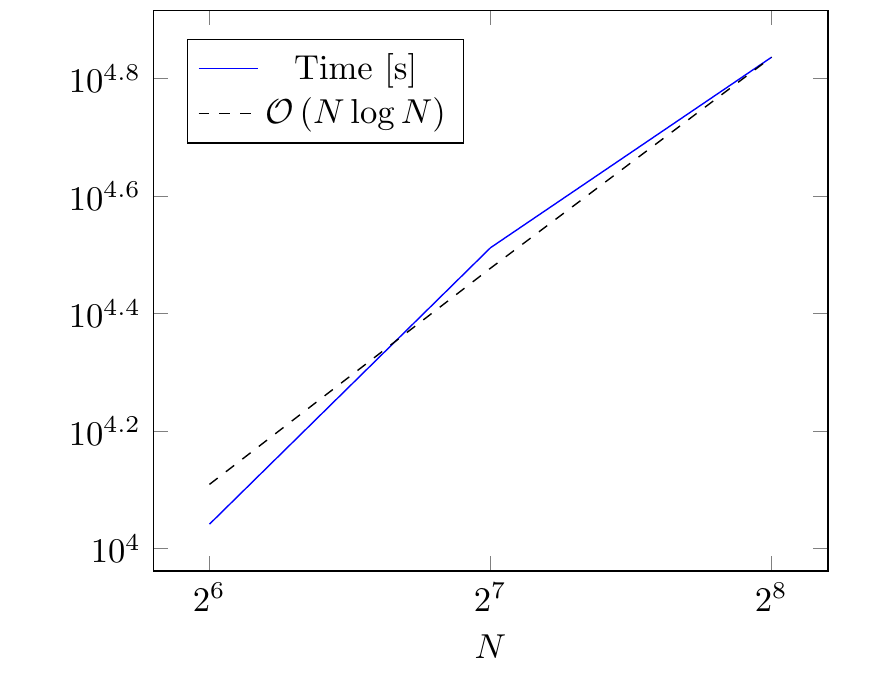}
        \caption{}
        \label{subfig:TimeScaleQuantum}
    \end{subfigure}
    \caption{Size in gates (\subref{subfig:GateScale}) of quantum circuits, and corresponding time to compute (\subref{subfig:TimeScaleQuantum}).
             $\mathcal{O}\left(N\log N\right)$ line shown for comparison.}
    \label{fig:QuantumScaling}
\end{center}
\end{figure*}
Fig. \ref{fig:TimeScaleClassical} shows the time needed to apply the inverse preconditioner classically using GMRES for various $N$.
For the preconditioning scheme used, matrices contained ~5 elements per row, with no significant dependence on $N$, and hence $N_{nz}=\mathcal{O}\left(N\right)$.
We see that for large matrices, the time required increases roughly as $\mathcal{O}\left(N^3\right)$, showing that no advantage whatsoever is gained from the sparsity of the initial matrix.
For the quantum procedure, Fig. \ref{fig:QuantumScaling} shows the size in gates of the final preconditioner inversion circuits, as well as the time taken to compute these circuits.
We see that the expected $\mathcal{O}\left(N\log\left(N\right)\right)$ scaling is obtained, and therefore the quantum procedure is, asymptotically, more efficient than the classical procedure.

\begin{figure}
\begin{center}
    \includegraphics[width=\columnwidth]{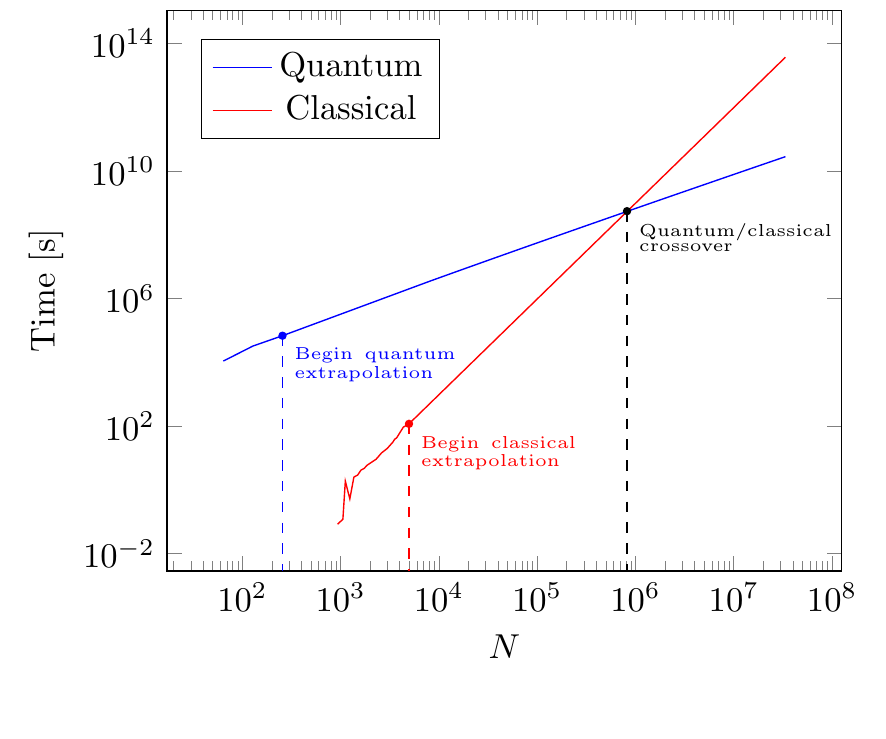}
    \caption{Simple extrapolation from computed data sets to estimate quantum/classical efficiency crossover.
            $\mathcal{O}\left(N\log N\right)$ and $\mathcal{O}\left(N^3\right)$ trend lines are used for the quantum and classical extrapolation, respectively.}
    \label{fig:Extrapolation}
\end{center}
\end{figure}
It is worth noting that the quantum procedure takes a very long time to produce a solution circuit for even modest problem sizes, 
limiting practical applicability to extremely large problems where the improved scaling can overcome the initial time difference.
Fig. \ref{fig:Extrapolation} gives a first-order extrapolation of our results to provide a rough prediction of the point where the quantum procedure can be expected to outperform a classical system.
We predict this to hold true for problems of more than roughly $8\times10^5$ unknowns, which would certainly indicate a relatively sophisticated problem.

Unfortunately, we are not able to analyze the time complexity of the solution circuit execution due to the limitations of available hardware.
The total number of qubits required for a system of size $N$ is
\begin{equation}
    2n + 2\left(n-1\right) + 2 + n_p + 1 = 4n + n_p + 1,
\end{equation}
accounting for the base registers and their ancillae, the control registers, the phase register 
(which could also require its own work register for controls, but we consider it to have constant size and therefore omit this extra dependence), and the HHL ancilla.
Even with a trivial single-qubit phase estimation, a system of 26 qubits would be required to invert our smallest preconditioner ($N=64$).
Since the largest system we can access provides only $7$ qubits, this calculation is quite beyond the capabilities of available hardware.
Note that, by this calculation, $82$ qubits would be required at our estimated crossover point of quantum/classical efficiency (using $n=20$ as the smallest sufficient power of two).

\section{Conclusion}
By combining the known HHL algorithm with a unitary adopted from quantum walk research, 
we have developed a method for the solution of any well-conditioned matrix equation which is suitable for direct implementation on quantum hardware.
The method has demonstrated $\mathcal{O}\left(N_{nz}\log\left(N\right)\right)$ complexity in time and gate count,
and for sparse matrices is expected to outperform classical solvers for problems of $N\gtrapprox8\times10^{5}$ unknowns.

While the method itself is suitable for practical implementation, available quantum systems are far too small and unreliable to support the analysis of any nontrivial problem.
This situation is expected to improve rapidly, however, with IBM projecting the release of a 1,000 qubit system by the end of 2023\cite{Cho}.
They also anticipate a dramatic decrease in errors\cite{Council}.
Google has also indicated plans to develop a commercial-grade system by 2029\cite{Castellanos}.

\appendices

\section{Reflection Operators}\label{sec:Reflectors}
\begin{figure}
\begin{center}
    \includegraphics[width=\columnwidth]{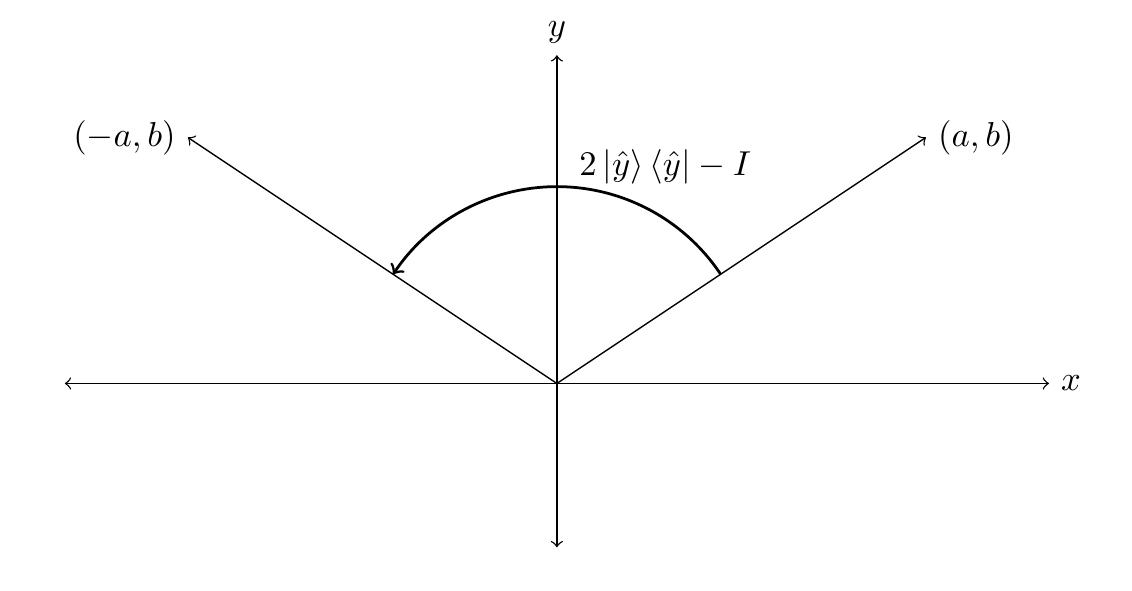}
    \caption{Two-dimensional reflection about the $y$-axis.
             }
    \label{fig:Reflection}
\end{center}
\end{figure}

Here we briefly describe the nature of reflection operators.
In particular, we consider operators of the form $2\ket{v}\bra{v}-I$, where $\ket{v}$ is some normalized vector.
This operator, when applied to another vector $\ket{w}$ of the same dimension, reflects $\ket{w}$ about $\ket{v}$.
As an initial example, consider the two-dimensional case where $\ket{v}$ is the unit vector along the $y$ axis, $\ket{\hat{y}}$, 
and $\ket{w}=a\ket{\hat{x}}+b\ket{\hat{y}}$ is arbitrary.
Then the effect of the reflection operator is as follows:
\begin{align}
    &\left(2\ket{\hat{y}}\bra{\hat{y}}-I\right)\left(a\ket{\hat{x}}+b\ket{\hat{y}}\right)
    \nonumber \\
    &=
    \left(2a\ket{\hat{y}}\braket{\hat{y}|\hat{x}} + 2b\ket{\hat{y}}\braket{\hat{y}|\hat{y}}\right) - \left(a\ket{\hat{x}}+b\ket{\hat{y}}\right)
    \nonumber \\
    &= 2b\ket{\hat{y}} - \left(a\ket{\hat{x}}+b\ket{\hat{y}}\right)
    \nonumber \\
    &= -a\ket{\hat{x}} + b\ket{\hat{y}}.
\end{align}
That is, the $x$ component of $\ket{w}$ is negated, while the $y$ component remains unchanged.
Fig. \ref{fig:Reflection} illustrates this behavior, and emphasizes the reflection of the vector.

In general, the operand vector $\ket{w}$ consists of components both perpendicular and parallel to $\ket{v}$.
Thus the vector can be written as $\ket{w}=a\ket{w_{\parallel}}+b\ket{w_{\perp}}$, and the action of the reflector is
\begin{align}
    &\left(2\ket{v}\bra{v}-I\right)\left(a\ket{w_{\parallel}}+b\ket{w_{\perp}}\right)
    \nonumber \\
    &=
    \left(2a\ket{v}\braket{v|w_{\parallel}} + 2b\ket{v}\braket{v|w_{\perp}}\right) - \left(a\ket{w_{\parallel}}+b\ket{w_{\perp}}\right)
    \nonumber \\
    &= 2a\ket{w_{\parallel}} - \left(a\ket{w_{\parallel}} + b\ket{w_{\perp}}\right)
    \nonumber \\
    &= a\ket{w_{\parallel}} - b\ket{w_{\perp}}.
\end{align}
Then the arbitrary reflector $2\ket{v}\bra{v}-I$ has the effect of negating the perpendicular component of the operand vector, while leaving the parallel component unchanged.

\section{Quantum Circuit Components}\label{sec:Components}
This section lists the individual operations which constitute each component of the divided circuit described in section \ref{ssec:Quantum}.
Operations are given in Qiskit-style syntax similar to that used in the program.
Operators are listed first, with the operand registers listed on the following line.
We assume that $C$, $X$, and $d$ have been defined.
Note that each of these components also includes the operations necessary to initialize it to the ideal state resulting from the precise application of all previous components.
These operations, applied in sequence to an appropriately initialized system, constitute the solution procedure for the relevant system.

\begin{Verbatim}[numbers=left,xleftmargin=5mm,fontsize=\tiny]
# initial T0
HGate()
    reg_r2

# QPE
HGate()
    [reg_phase[0]]
HGate()
    [reg_phase[1]]

HGate().control(2,None,'01')
    [reg_phase[0]]+[reg_r1a[0]]+[reg_r2[0]]
XGate().control(2,None,'11')
    [reg_phase[0]]+[reg_r1a[0]]+[reg_r2a[0]]
ZGate().control(1,None,'1')
    [reg_phase[0]]+[reg_r2[0]]
XGate().control(1,None,'1')
    [reg_phase[0]]+[reg_r2[0]]
ZGate().control(1,None,'1')
    [reg_phase[0]]+[reg_r2[0]]
XGate().control(1,None,'1')
    [reg_phase[0]]+[reg_r2[0]]
XGate().control(2,None,'01')
    [reg_phase[0]]+reg_r2[:]+reg_r2a[:]
ZGate().control(2,None,'01')
    [reg_phase[0]]+reg_r2[:]+reg_r2a[:]
XGate().control(2,None,'01')
    [reg_phase[0]]+reg_r2[:]+reg_r2a[:]
HGate().control(2,None,'01')
    [reg_phase[0]]+[reg_r1a[0]]+[reg_r2[0]]
XGate().control(2,None,'11')
    [reg_phase[0]]+[reg_r1a[0]]+[reg_r2a[0]]
SwapGate().control(1,None,'1')
    [reg_phase[0]]+reg_r1[:]+reg_r2[:]
SwapGate().control(1,None,'1')
    [reg_phase[0]]+reg_r1a[:]+reg_r2a[:]
SGate().control(1,None,'1')
    [reg_phase[0]]+[reg_r1[0]]
XGate().control(1,None,'1')
    [reg_phase[0]]+[reg_r1[0]]
SGate().control(1,None,'1')
    [reg_phase[0]]+[reg_r1[0]]
XGate().control(1,None,'1')
    [reg_phase[0]]+[reg_r1[0]]

HGate().control(2,None,'01')
    [reg_phase[1]]+[reg_r1a[0]]+[reg_r2[0]]
XGate().control(2,None,'11')
    [reg_phase[1]]+[reg_r1a[0]]+[reg_r2a[0]]
ZGate().control(1,None,'1')
    [reg_phase[1]]+[reg_r2[0]]
XGate().control(1,None,'1')
    [reg_phase[1]]+[reg_r2[0]]
ZGate().control(1,None,'1')
    [reg_phase[1]]+[reg_r2[0]]
XGate().control(1,None,'1')
    [reg_phase[1]]+[reg_r2[0]]
XGate().control(2,None,'01')
    [reg_phase[1]]+reg_r2[:]+reg_r2a[:]
ZGate().control(2,None,'01')
    [reg_phase[1]]+reg_r2[:]+reg_r2a[:]
XGate().control(2,None,'01')
    [reg_phase[1]]+reg_r2[:]+reg_r2a[:]
HGate().control(2,None,'01')
    [reg_phase[1]]+[reg_r1a[0]]+[reg_r2[0]]
XGate().control(2,None,'11')
    [reg_phase[1]]+[reg_r1a[0]]+[reg_r2a[0]]
SwapGate().control(1,None,'1')
    [reg_phase[1]]+reg_r1[:]+reg_r2[:]
SwapGate().control(1,None,'1')
    [reg_phase[1]]+reg_r1a[:]+reg_r2a[:]
SGate().control(1,None,'1')
    [reg_phase[1]]+[reg_r1[0]]
XGate().control(1,None,'1')
    [reg_phase[1]]+[reg_r1[0]]
SGate().control(1,None,'1')
    [reg_phase[1]]+[reg_r1[0]]
XGate().control(1,None,'1')
    [reg_phase[1]]+[reg_r1[0]]

HGate().control(2,None,'01')
    [reg_phase[1]]+[reg_r1a[0]]+[reg_r2[0]]
XGate().control(2,None,'11')
    [reg_phase[1]]+[reg_r1a[0]]+[reg_r2a[0]]
ZGate().control(1,None,'1')
    [reg_phase[1]]+[reg_r2[0]]
XGate().control(1,None,'1')
    [reg_phase[1]]+[reg_r2[0]]
ZGate().control(1,None,'1')
    [reg_phase[1]]+[reg_r2[0]]
XGate().control(1,None,'1')
    [reg_phase[1]]+[reg_r2[0]]
XGate().control(2,None,'01')
    [reg_phase[1]]+reg_r2[:]+reg_r2a[:]
ZGate().control(2,None,'01')
    [reg_phase[1]]+reg_r2[:]+reg_r2a[:]
XGate().control(2,None,'01')
    [reg_phase[1]]+reg_r2[:]+reg_r2a[:]
HGate().control(2,None,'01')
    [reg_phase[1]]+[reg_r1a[0]]+[reg_r2[0]]
XGate().control(2,None,'11')
    [reg_phase[1]]+[reg_r1a[0]]+[reg_r2a[0]]
SwapGate().control(1,None,'1')
    [reg_phase[1]]+reg_r1[:]+reg_r2[:]
SwapGate().control(1,None,'1')
    [reg_phase[1]]+reg_r1a[:]+reg_r2a[:]
SGate().control(1,None,'1')
    [reg_phase[1]]+[reg_r1[0]]
XGate().control(1,None,'1')
    [reg_phase[1]]+[reg_r1[0]]
SGate().control(1,None,'1')
    [reg_phase[1]]+[reg_r1[0]]
XGate().control(1,None,'1')
    [reg_phase[1]]+[reg_r1[0]]

SwapGate()
    reg_phase
HGate()
    [reg_phase[0]]
CZGate().power(0.5).inverse()
    reg_phase
HGate()
    [reg_phase[1]]

# HHL ancilla rotation
RYGate(2.0*math.acos(C/(-d))).control(nq_phase,None,'00')
    reg_phase[:]+reg_a_hhl[:]
RYGate(2.0*math.acos(C/(X-d))).control(nq_phase,None,'01')
    reg_phase[:]+reg_a_hhl[:]
RYGate(2.0*math.acos(C/(-d))).control(nq_phase,None,'10')
    reg_phase[:]+reg_a_hhl[:]
RYGate(2.0*math.acos(C/(-X-d))).control(nq_phase,None,'11')
    reg_phase[:]+reg_a_hhl[:]

# inverse QPE
HGate()
    [reg_phase[1]]
CZGate().power(0.5)
    reg_phase
HGate()
    [reg_phase[0]]
SwapGate()
    reg_phase

XGate().control(1,None,'1')
    [reg_phase[1]]+[reg_r1[0]]
SGate().inverse().control(1,None,'1')
    [reg_phase[1]]+[reg_r1[0]]
XGate().control(1,None,'1')
    [reg_phase[1]]+[reg_r1[0]]
SGate().inverse().control(1,None,'1')
    [reg_phase[1]]+[reg_r1[0]]
SwapGate().control(1,None,'1')
    [reg_phase[1]]+reg_r1a[:]+reg_r2a[:]
SwapGate().control(1,None,'1')
    [reg_phase[1]]+reg_r1[:]+reg_r2[:]
XGate().control(2,None,'11')
    [reg_phase[1]]+[reg_r1a[0]]+[reg_r2a[0]]
HGate().control(2,None,'01')
    [reg_phase[1]]+[reg_r1a[0]]+[reg_r2[0]]
XGate().control(2,None,'01')
    [reg_phase[1]]+reg_r2[:]+reg_r2a[:]
ZGate().control(2,None,'01')
    [reg_phase[1]]+reg_r2[:]+reg_r2a[:]
XGate().control(2,None,'01')
    [reg_phase[1]]+reg_r2[:]+reg_r2a[:]
XGate().control(1,None,'1')
    [reg_phase[1]]+[reg_r2[0]]
ZGate().control(1,None,'1')
    [reg_phase[1]]+[reg_r2[0]]
XGate().control(1,None,'1')
    [reg_phase[1]]+[reg_r2[0]]
ZGate().control(1,None,'1')
    [reg_phase[1]]+[reg_r2[0]]
XGate().control(2,None,'11')
    [reg_phase[1]]+[reg_r1a[0]]+[reg_r2a[0]]
HGate().control(2,None,'01')
    [reg_phase[1]]+[reg_r1a[0]]+[reg_r2[0]]

XGate().control(1,None,'1')
    [reg_phase[1]]+[reg_r1[0]]
SGate().inverse().control(1,None,'1')
    [reg_phase[1]]+[reg_r1[0]]
XGate().control(1,None,'1')
    [reg_phase[1]]+[reg_r1[0]]
SGate().inverse().control(1,None,'1')
    [reg_phase[1]]+[reg_r1[0]]
SwapGate().control(1,None,'1')
    [reg_phase[1]]+reg_r1a[:]+reg_r2a[:]
SwapGate().control(1,None,'1')
    [reg_phase[1]]+reg_r1[:]+reg_r2[:]
XGate().control(2,None,'11')
    [reg_phase[1]]+[reg_r1a[0]]+[reg_r2a[0]]
HGate().control(2,None,'01')
    [reg_phase[1]]+[reg_r1a[0]]+[reg_r2[0]]
XGate().control(2,None,'01')
    [reg_phase[1]]+reg_r2[:]+reg_r2a[:]
ZGate().control(2,None,'01')
    [reg_phase[1]]+reg_r2[:]+reg_r2a[:]
XGate().control(2,None,'01')
    [reg_phase[1]]+reg_r2[:]+reg_r2a[:]
XGate().control(1,None,'1')
    [reg_phase[1]]+[reg_r2[0]]
ZGate().control(1,None,'1')
    [reg_phase[1]]+[reg_r2[0]]
XGate().control(1,None,'1')
    [reg_phase[1]]+[reg_r2[0]]
ZGate().control(1,None,'1')
    [reg_phase[1]]+[reg_r2[0]]
XGate().control(2,None,'11')
    [reg_phase[1]]+[reg_r1a[0]]+[reg_r2a[0]]
HGate().control(2,None,'01')
    [reg_phase[1]]+[reg_r1a[0]]+[reg_r2[0]]

XGate().control(1,None,'1')
    [reg_phase[0]]+[reg_r1[0]]
SGate().inverse().control(1,None,'1')
    [reg_phase[0]]+[reg_r1[0]]
XGate().control(1,None,'1')
    [reg_phase[0]]+[reg_r1[0]]
SGate().inverse().control(1,None,'1')
    [reg_phase[0]]+[reg_r1[0]]
SwapGate().control(1,None,'1')
    [reg_phase[0]]+reg_r1a[:]+reg_r2a[:]
SwapGate().control(1,None,'1')
    [reg_phase[0]]+reg_r1[:]+reg_r2[:]
XGate().control(2,None,'11')
    [reg_phase[0]]+[reg_r1a[0]]+[reg_r2a[0]]
HGate().control(2,None,'01')
    [reg_phase[0]]+[reg_r1a[0]]+[reg_r2[0]]
XGate().control(2,None,'01')
    [reg_phase[0]]+reg_r2[:]+reg_r2a[:]
ZGate().control(2,None,'01')
    [reg_phase[0]]+reg_r2[:]+reg_r2a[:]
XGate().control(2,None,'01')
    [reg_phase[0]]+reg_r2[:]+reg_r2a[:]
XGate().control(1,None,'1')
    [reg_phase[0]]+[reg_r2[0]]
ZGate().control(1,None,'1')
    [reg_phase[0]]+[reg_r2[0]]
XGate().control(1,None,'1')
    [reg_phase[0]]+[reg_r2[0]]
ZGate().control(1,None,'1')
    [reg_phase[0]]+[reg_r2[0]]
XGate().control(2,None,'11')
    [reg_phase[0]]+[reg_r1a[0]]+[reg_r2a[0]]
HGate().control(2,None,'01')
    [reg_phase[0]]+[reg_r1a[0]]+[reg_r2[0]]

HGate()
    [reg_phase[1]]
HGate()
    [reg_phase[0]]

# inverse T0
HGate()
    reg_r2
\end{Verbatim}

\section*{Acknowledgment}
We acknowledge the use of IBM Quantum services for this work.
The views expressed are those of the authors, and do not reflect the official policy or position of IBM or the IBM Quantum team.

\end{document}